\documentclass[nofootinbib,amsmath,amssymb,aps,floatfix,notitlepage,
superscriptaddress,
longbibliography]{revtex4-2}
\pdfoutput=1

\usepackage{graphics}
\usepackage{graphicx}
\usepackage{color}

\usepackage{enumitem}

\usepackage{amssymb}

\usepackage{hyperref}\hypersetup{colorlinks=true,linktoc=all,}

\def\hhref#1{\href{http://arxiv.org/abs/#1}{arXiv:#1}} 

\begin{document}
\title{Conformal and Uniformizing Maps in Borel Analysis}

\author{Ovidiu Costin}
\affiliation{Department of Mathematics, The Ohio State University, Columbus OH 43210-1174, USA}
\author{Gerald V. Dunne}
\affiliation{Department of Physics, University of Connecticut, Storrs CT 06269-3046, USA}

%
\begin{abstract} 
Perturbative expansions in physical applications are generically divergent, and their physical content can be studied using Borel analysis. Given just a finite number of terms of such an expansion, this input data can be analyzed in different ways, leading to vastly different precision for the extrapolation of the expansion parameter away from its original asymptotic regime. Here we describe how conformal maps and uniformizing maps can be used, in conjunction with Pad\'e approximants, to increase the precision of the information that can be extracted from a finite amount of perturbative input data. We also summarize results from the physical interpretation of Pad\'e approximations in terms of electrostatic potential theory.
%
\end{abstract}

\maketitle
\section{Introduction}
\label{intro}
Perturbation theory is generically divergent, but in principle it contains a wealth of physical information also about non-perturbative effects \cite{leguillou}. However, in difficult problems it is often challenging to compute many orders of a perturbative expansion. This raises the mathematical question of how best to extract as much information as possible from a finite number of terms of an expansion. The first main conclusion is that it is more effective to map the problem to the Borel plane, converting the divergent series to a convergent Borel transform function, dividing out the generic leading factorial growth. This simple step {\it smoothens} the problem, enabling the use of a vast array of powerful tools of complex analysis. At a deeper level one can use the full machinery of resurgent asymptotics \cite{ecalle,costin-book} to probe the system in all parametric regions. The question now becomes: given a finite number of terms of an expansion of a Borel transform function $\mathcal{B}(p)$, what can we learn about its analytic structure in the complex $p$ plane? Singularities of $\mathcal{B}(p)$ have special physical significance, as they are directly related to non-perturbative physics and with associated Stokes phenomena.

Thus we can summarize our motivation as follows: given a {\it finite number of terms} in the expansion of a Borel function $\mathcal B(p)$ about some point (say, $p=0$), we ask what is the optimal procedure, and what are effective near-optimal approximations to answer the following questions:
\begin{enumerate}
\item
{\it Where} are the singularities $p_0$ of $\mathcal B(p)$ in the complex $p$ plane? This might refer, for example, to locating a phase transition.
\item
What is the {\it nature} (exponent $\alpha$)  of each singularity? $\mathcal B(p)\sim (p-p_0)^\alpha$. This is relevant for determining critical exponents.
\item
What is the {\it coefficient} of the leading singularity (or leading singularities)? $\mathcal B(p)\sim \mathcal S (p-p_0)^\alpha$. This relates to the determination of Stokes constants.

\item
Can one extrapolate from one Riemann sheet to the next? This is also important for the study of phase transitions.

\end{enumerate}

Many ingenious {\it ad hoc} methods have been developed for different mathematical and physical applications. Here we describe recent work aimed at developing a systematic mathematical framework to find optimal methods to solve these problems  \cite{Costin:2019xql,Costin:2020hwg,Costin:2020pcj}. We refer to this as ``inverse approximation theory'', because the task is to learn as much as possible about some function, based on partial information about it, rather than finding efficient approximations for a given function.  For the wide class of resurgent functions, which arise frequently in applications and are expected to suffice for all natural problems, new practical approximation methods can achieve near-optimal results, especially in the vicinity of the singularities.

In this paper we restrict our attention to Borel analysis, but note that there are also interesting applications to certain problems where the physical function of interest is itself convergent, but we wish to learn about its radius of convergence (i.e., its singularities). A simple example is the Ising model, where the free energy has a finite radius of convergence, and we might wish to determine the behavior in the vicinity of the critical temperature by extrapolating from an expansion at either low or high temperature.

The key mathematical tools described here are Borel summation \cite{costin-book}, Pad\'e approximants \cite{baker}, conformal maps  and uniformizing maps \cite{bateman,Nehari}. These are of course well-studied methods, which have been used empirically in these kinds of investigations \cite{zj,caliceti,caprini}. Here we seek to explain how and why these methods work, and to compare their precision in quantifiable ways. The main message is that by combining them in various different ways, we can obtain dramatically improved precision in exploring the motivating questions listed above. In situations where we have physical input, even if it is conjectural and/or approximate,  concerning the analytic structure of the function under investigation (for example, unitarity or a dispersion relation or a particular symmetry), this information can be used to advantage, and can also be refined iteratively.

\section{Physics of Pad\'e Approximation: Electrostatic Potential Theory}
\label{sec:pade}

Pad\'e approximation is a versatile tool for analytic continuation of a function for which only a finite number of expansion coefficients is known \cite{baker,bender}.  Here we summarize some important results from the physical interpretation of Pad\'e approximation in terms of electrostatic potential theory \cite{Stahl,Saff,Costin:2020pcj}. 

The $[M, N]$ Pad\'e approximant of $B(p)$  at $p=0$ is the unique rational function $ P_M(p)/Q_N(p)$, with $P_M$ a polynomial of degree at most $M$, and $Q_N$ a polynomial of degree at most $N$, for which
\begin{equation}
  \label{eq:defPade}
  B(p)-\frac{P_M(p)}{Q_N(p)}=\mathcal{O}\left(p^{M+N+1}\right),\ \ \ p\to 0
\end{equation}
If we normalize $Q_N(0)=1$, the Pad\'e polynomials $P_M(p)$ and $Q_N(p)$ are also unique, and can be calculated  algorithmically from the (truncated) Maclaurin series of $B(p)$.
\\

\noindent{\bf Comments:} 
\begin{itemize}[label=$\circ$]
\item
In many applications, diagonal and near-diagonal Pad\'e approximants are the most useful.

\item 
There is a deep connection between Pad\'e approximants and orthogonal polynomials (and their associated large-order Szeg\"o asymptotics \cite{szego,Yattselev}) from the fundamental fact that the Pad\'e polynomials $P_M$ and $Q_N$ satisfy a three-term recursion relation \cite{baker,bender}. 

\item  In special cases, the convergence of Pad\'e approximants is uniform on compact sets. This is the case for Riesz-Markov functions (i.e., $B(p)=\int_a^b d\mu(\zeta)/(\zeta-p)$,  with $\mu$ a positive measure). See \cite{Damanik-Simon} and references therein. While these occur  in certain applications, many functions of interest are not  Riesz-Markov. For example, a common situation in applications, discussed in more detail in Section \ref{sec:pair} below, is when $B(p)$ has two complex conjugate singularities. Pad\'e produces curved arcs of poles (see Figure \ref{fig:pb-pair}) which do not relate to the properties of the function.

\item
 In fact, even for single-valued functions, uniform convergence of {\em some} diagonal Pad\'e subsequence to general meromorphic functions (the Baker-Gammel-Wills conjecture  \cite{b-g-w}) does not hold  \cite{Lubinsky}.
   Pointwise convergence is prevented by the phenomenon  of spurious  poles, Froissart doublets \cite{froissart}: ``random'' pairs of a pole and a nearby zero,  unrelated to the function they approximate.
The ultimate source of these spurious poles is the fact that the associated polynomials are orthogonal on complex arcs,  without a bona-fide Hilbert space structure \cite{Stahl}.

\end{itemize}

\subsection{Potential Theory and Physical Interpretation of Pad\'e Approximants}
\label{sec:potential}

Electrostatic potential theory provides a remarkable and intuitively useful physical interpretation of the Pad\'e domain of convergence and of the location of Pad\'e poles \cite{Stahl,Saff,Costin:2020pcj}. For this interpretation, it is useful to invert ($p\to 1/p$) to move the point of analyticity from $p=0$ to $p=\infty$.
Being  {\it rational approximations}, Pad\'e approximants can only converge in some domain $\mathcal{D}$ of single-valuedness of the associated function $B(p)$. Furthermore, in general  Pad\'e only converges in a weak sense, namely ``in capacity''. This means the following.
  Take any set $\mathcal{D}'$ of single-valuedness of $B(p)$, with boundary   $E'=\partial \mathcal{D}'$. Thinking of $E'$ as an electrical conductor we place a unit charge on $E'$, and normalize the electrostatic potential $V(x,y)=V(p), p=x+iy$ (always constant along a conductor) by $V(E')=0$. Then the electrostatic capacitance of $E'$ is cap$(E')= 1/V(\infty)$.

The fundamental result \cite{Stahl} is that the  boundary $E=\partial \mathcal{D}$ of the domain of convergence $\mathcal{D}$ of Pad\'e is obtained by deforming the shape  of the conductor $E'$ (while keeping the singularity locations fixed)  until the logarithmic  capacity is minimized.
Furthermore, the equilibrium measure $\mu$  on $E$ is the equilibrium density of charges on $E$.
 For points $p\in \mathcal{D}$, the potential is related to the Green's function $g_{\mathcal{D}}(p)$ as: $e^{-g_{\mathcal{D}}(p)}
 =e^{-V(p)}$.
  \\
  
  \noindent{\bf Comments:}

\begin{itemize}[label=$\circ$]

\item

Pad\'e   ``constructs'' the  maximal domain $\mathcal{D}$ of single-valuedness in which they converge. The rate of convergence near the point of expansion is given by the inverse of the (minimal)
capacity and,  in a precise sense,  it is provably optimal in the class
of rational approximations  \cite{Costin:2020pcj}.

\item
Pad\'e represents actual poles of $B(p)$ by poles, and branch points by lines (either straight or curved arcs) of poles accumulating to the branch point.
If $B(p)$ has only isolated singularities on the universal cover of $\hat{\mathbb C}$ with finitely many punctures, then the boundary of single-valuedness,
$\partial \mathcal{D}$, is a set of piecewise analytic arcs joining the branch points of $B(p)$, and some accessory points associated with junctions of these analytic arcs. See for example Figures \ref{fig:pb-pair}  and \ref{fig:probe}. 
The pole density converges in capacity to the equilibrium measure along the arcs, and this density is infinite at the actual branch points, resulting in an accumulation of poles there.

  \item If $\mathcal{D}$ is simply connected, then $|e^{-g_{\mathcal{D}}}|=|\psi_{\infty}|$, where $\psi_\infty$ is a conformal map from $\mathcal{D}$ to the interior of the unit disk,  normalized with $\psi_\infty(\infty)=0$. Thus Pad\'e effectively ``creates its own conformal map'' of a single-valuedness domain for $B(p)$. This map can be extracted
    in the  $N\to \infty$ limit from the harmonic function $|e^{-g_{\mathcal{D}}}|$, obtained  by taking the $N$-th root of the convergence rate.

\item

However, this highlights an inherent drawback of the Pad\'e construction:  points of interest of $B(p)$ may be hidden on the boundary of convergence.  For example, true Borel singularities of $B(p)$ may be obscured by the lines of Pad\'e poles of the minimal capacitor.  Nevertheless, the physical intuition behind this potential theory interpretation of Pad\'e approximation enables several simple methods to reveal such ``hidden'' singularities using conformal maps or probe singularities. See for example the discussion in Section \ref{sec:multi} below, and Figures \ref{fig:hidden}  and \ref{fig:probe}.

\end{itemize}

\section{Borel Transform Functions With Branch Points}

The generic singularity of a Borel transform in physical applications is a branch point. Often there is a {\it dominant} Borel singularity, associated with the leading divergence of the associated asymptotic series. Cases with multiple Borel singularities are discussed below in Sections \ref{sec:pair} and  \ref{sec:multi}, in which case there can be interesting interference effects leading to a richer structure of divergent series.

\subsection{One Branch Point Borel Singularity}
\label{sec:one-borel}

In many physical applications we encounter a formal asymptotic series
\begin{eqnarray}
f(x)
\sim \sum_{n=0}^\infty \frac{c_n}{x^{n+1}}
\qquad, \quad x\to+\infty
\label{eq:series}
\end{eqnarray}
where the leading rate of growth of the coefficients $c_n$  has the ``power times factorial'' form
\cite{leguillou}
\begin{eqnarray}
c_n \sim A^n\,  \Gamma(B\, n+C)\qquad, \quad n\to +\infty
\label{eq:bwl}
\end{eqnarray}
Here the parameters $A$, $B$ and $C$ are constants. In treating such a problem, a natural approach is to apply Borel summation, leading to a Borel-Laplace integral representation of a function $f(x)$ with the asymptotic series (\ref{eq:series}):
\begin{eqnarray}
f(x)=\int_0^\infty dp\, e^{-p\, x} \mathcal B(p)
\label{eq:laplace}
\end{eqnarray}
This maps the problem of extrapolating $f(x)$ into the complex $x$ plane to the problem of understanding the analytic structure of the Borel transform $\mathcal{B}(p)$, particularly its singularities in the complex Borel $p$ plane.

For example, the following function (based on the incomplete gamma function)
\begin{eqnarray}
F(x; \alpha) \equiv x^{-1-\alpha} \, e^x \, \Gamma(1+\alpha, x)
=\int_0^\infty dp\, e^{-p\, x} (1+p)^\alpha 
\label{eq:i-gamma}
\end{eqnarray}
has an asymptotic expansion as $x\to+\infty$ of the form in (\ref{eq:series})-(\ref{eq:bwl}):
\begin{eqnarray}
F(x; \alpha)
\sim \frac{1}{\Gamma(-\alpha)} \sum_{n=0}^\infty (-1)^n  \frac{\Gamma(n-\alpha)}{x^{n+1}}
\qquad, \quad x\to+\infty
\label{eq:inc-gamma2}
\end{eqnarray}
$F(x; \alpha)$ 
has non-trivial 
analytic continuation properties in the complex $x$ plane which are encoded in the analytic structure of the Borel transform function $\mathcal B(p)=(1+p)^\alpha$.

Another example is  the function\footnote{For example, the Airy function is ${\rm Ai}(x)=\frac{2 x^{5/4}}{3\sqrt{\pi}}e^{-\frac{2}{3} x^{3/2}} F\left(\frac{4}{3} x^{3/2};\frac{1}{6}, \frac{5}{6}, 1\right)$, and the Whittaker function is $W_{\mu,\nu}(x)=x^{1+\mu}\, e^{-x/2} F\left(x; \frac{1}{2}+\nu-\mu, \frac{1}{2}-\nu-\mu, 1\right)$.}
\begin{eqnarray}
F(x; a, b, c)\equiv \int_0^\infty dp\, e^{-p\, x} ~_2F_1(a, b, c; -p)
\label{eq:f-hyper}
\end{eqnarray}
which has an asymptotic expansion as $x\to+\infty$
\begin{eqnarray}
F(x; a, b, c) \sim \frac{\Gamma(c)}{\Gamma(a)\, \Gamma(b)}\sum_{n=0}^\infty (-1)^n \frac{\Gamma(n+a)\, \Gamma(n+b)}{\Gamma(n+c)}\frac{1}{x^{n+1}}\qquad, \quad x\to +\infty
\label{eq:f-hyper-exp}
\end{eqnarray}
The expansion coefficients have leading large order growth as in (\ref{eq:bwl}), but now with further subleading power-law corrections:
\begin{eqnarray}
c_n\sim  \frac{\Gamma(c)}{\Gamma(a)\, \Gamma(b)}\,(-1)^n\, \Gamma(n+a+b-c)\left[1-\frac{(a-c)(b-c)}{n}+O\left(\frac{1}{n^2}\right) \right]
\label{eq:hyper-cn}
\end{eqnarray}
The Borel transform functions  in (\ref{eq:i-gamma}) and (\ref{eq:f-hyper}), $\mathcal B(p)=(1+p)^\alpha$ and $\mathcal B(p)=~_2F_1(a, b, c; -p)$, respectively, have the common feature of possessing just one branch point, which we have normalized here to lie at $p=-1$. 

In the physically relevant situation in which we only know a {\it finite} number of coefficients $c_n$ of the asymptotic expansion (\ref{eq:series}), there are several different approaches to analyze the singularity structure of the associated Borel transform. Pad\'e approximants are a key tool in analytically continuing the truncated Borel transform beyond its radius of convergence, but they can be significantly improved by combining them with conformal and uniformizing maps.
\\

{\bf Pad\'e-Borel transform:} Pad\'e approximants provide remarkably accurate extrapolations and analytic continuations of truncated expansions \cite{baker,bender}. For a truncation of an {\it asymptotic} series it is generally better to apply a Pad\'e approximation in the Borel $p$ plane than in the original $x$ plane. Thus, we apply Pad\'e to the convergent Borel transform function rather than to the divergent large $x$ expansion (\ref{eq:series}). This procedure is referred to as Pad\'e-Borel ($\mathcal {PB}$). Note that Pad\'e is a non-linear operation, so it does not commute with the linear Borel transform (\ref{eq:laplace}). The higher precision of Pad\'e-Borel has been observed empirically and it is widely used  \cite{simon,zj,marino}. The improved precision has recently been  proven with explicit error bounds for the canonical example (\ref{eq:i-gamma}) \cite{Costin:2020hwg}, for which the exact $[N, N]$ diagonal Pad\'e-Borel transform is
\begin{eqnarray}
{\rm Pade}{\rm -}{\rm Borel:} \qquad {\mathcal {PB}}_{[N,N]}(p; \alpha)=\frac{P_N^{(\alpha,- \alpha)}\left(1+\frac{2}{p}\right)}{P_N^{(-\alpha, \alpha)}\left(1+\frac{2}{p}\right)}
\label{eq:pbp}
\end{eqnarray}
Here $P_N^{(\alpha, \beta)}$ is the $N^{\rm th}$ Jacobi polynomial. The large $N$ asymptotics of the Jacobi polynomials explains why $\mathcal {PB}$ yields such an improvement.
\\

{\bf Pad\'e-Conformal-Borel transform:} if there is a dominant Borel branch point singularity,\footnote{In the simple case where  the singularity is not a  branch point but a pole, Pad\'e-Borel  is of course optimal.} an even more accurate summation than $\mathcal {PB}$ is obtained by first making a conformal map from the cut Borel plane $p\in {\mathbb C}\setminus (-\infty, -1]$ to the interior of the unit disc $|z|<1$ using the conformal map:
\begin{eqnarray}
p=\frac{4z}{(1-z)^2}
\quad \longleftrightarrow \quad 
z=\frac{\sqrt{1+p}-1}{\sqrt{1+p}+1}
\label{eq:cmap}
\end{eqnarray}
and then making a Pad\'e approximant in the $z$ plane. 
This conformal map takes the singularity at $p=-1$ to $z=-1$, and $p=0$ to $z=0$, while $z=1$ corresponds to the point at infinity in the $p$ plane. The upper/lower edge of the cut $p\in (-\infty, -1]$ maps to upper/lower unit circle in the $z$ plane. For example, the points $p=-2\pm i\epsilon$ (as $\epsilon\to 0^+$) map to $z=\pm i$.
Conformal maps are frequently used in series analysis \cite{caprini,caliceti,zj}, and when combined with a subsequent Pad\'e approximant, there is a significant further gain in precision \cite{Costin:2020hwg}.\footnote{Without the additional Pad\'e approximation, the conformal map is only as effective as the $\mathcal{PB}$ approximation described above \cite{Costin:2020hwg}.} For example, 
 the $\mathcal{PB}$ result (\ref{eq:pbp}) is replaced by the closed-form Pad\'e-Conformal-Borel ($\mathcal {PCB}$) transform:
\begin{eqnarray}
{\rm Pade}{\rm -}{\rm Conformal}{\rm -}{\rm Borel:} \qquad {\mathcal {PCB}}_{[N,N]}(p; \alpha)=\frac{P_N^{(2\alpha,- 2\alpha)}\left(\frac{\sqrt{1+p}+1}{\sqrt{1+p}-1}\right)}{P_N^{(-2\alpha, 2\alpha)}\left(\frac{\sqrt{1+p}+1}{\sqrt{1+p}-1}\right)}
\label{eq:pcbp}
\end{eqnarray}
\\
\begin{figure}[htb]
\centerline{
\includegraphics[scale=0.7]{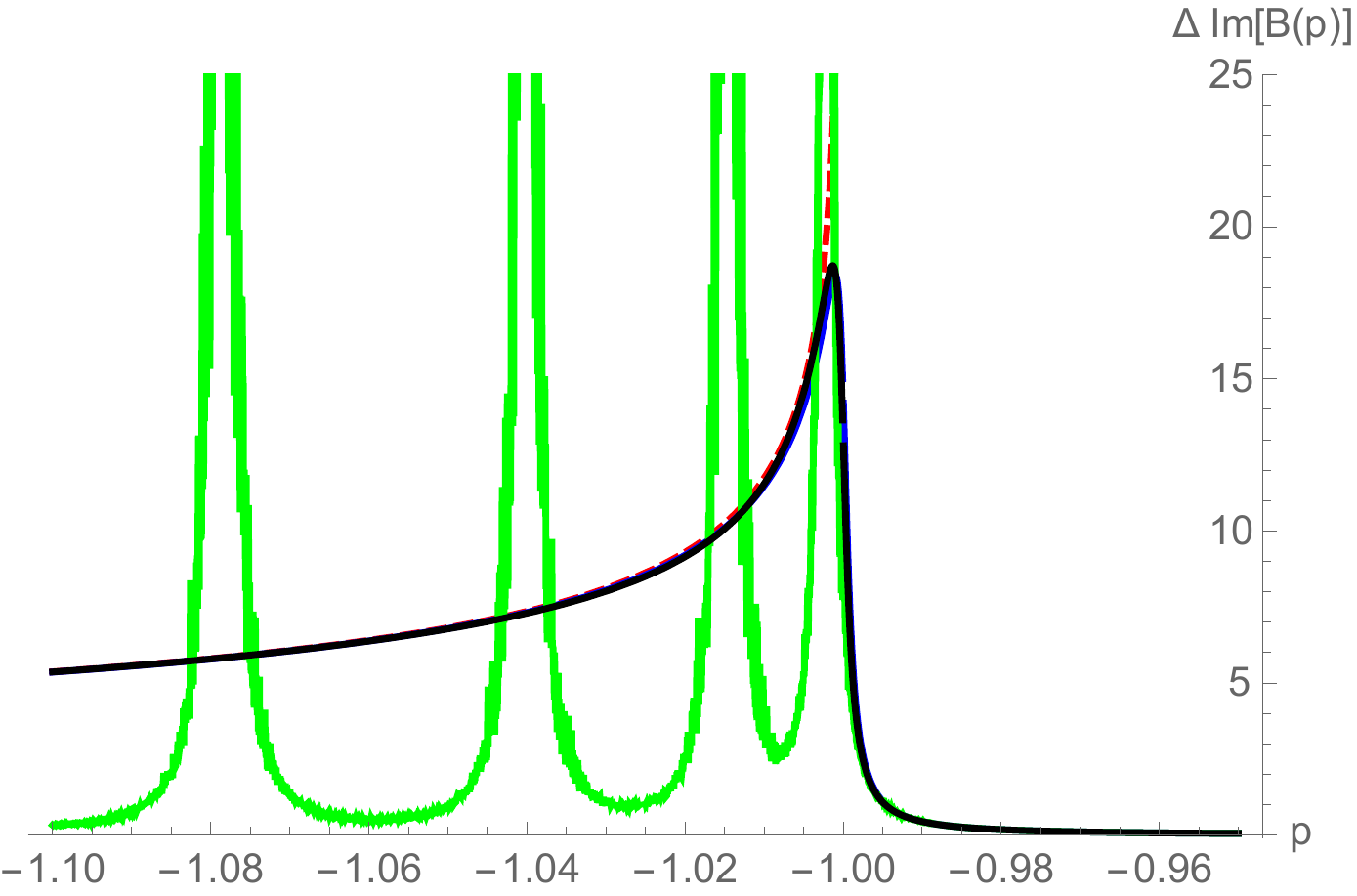}
}
\caption{Plot of the jump in the imaginary part of the hypergeometric Borel function 
$B(p)=~_2F_1\left(\frac{2}{3}, \frac{2}{3}, 1; -p\right)$, across the cut $p\in (-\infty, -1]$, based on just $10$ terms of a truncated asymptotic expansion. The blue and black curves are the jump for $p\pm 10^{-3} i$, just above and below the cut, for the exact function and its $\mathcal{PCB}$ and $\mathcal{PUB}$ approximations. These three curves are indistinguishable on this scale. The red dashed curve is the exact jump using properties of hypergeometric functions, while the green curve is the jump in the Pad\'e-Borel approximation. We see the effect of the Pad\'e poles along the cut. 
The jittery nature of the Pad\'e-Borel approximation can be improved by converting the Pad\'e polynomial ratio to a partial fraction expansion. }
\label{fig:hyper}
\end{figure}

\begin{figure}[htb]
\centerline{\includegraphics[scale=0.55]{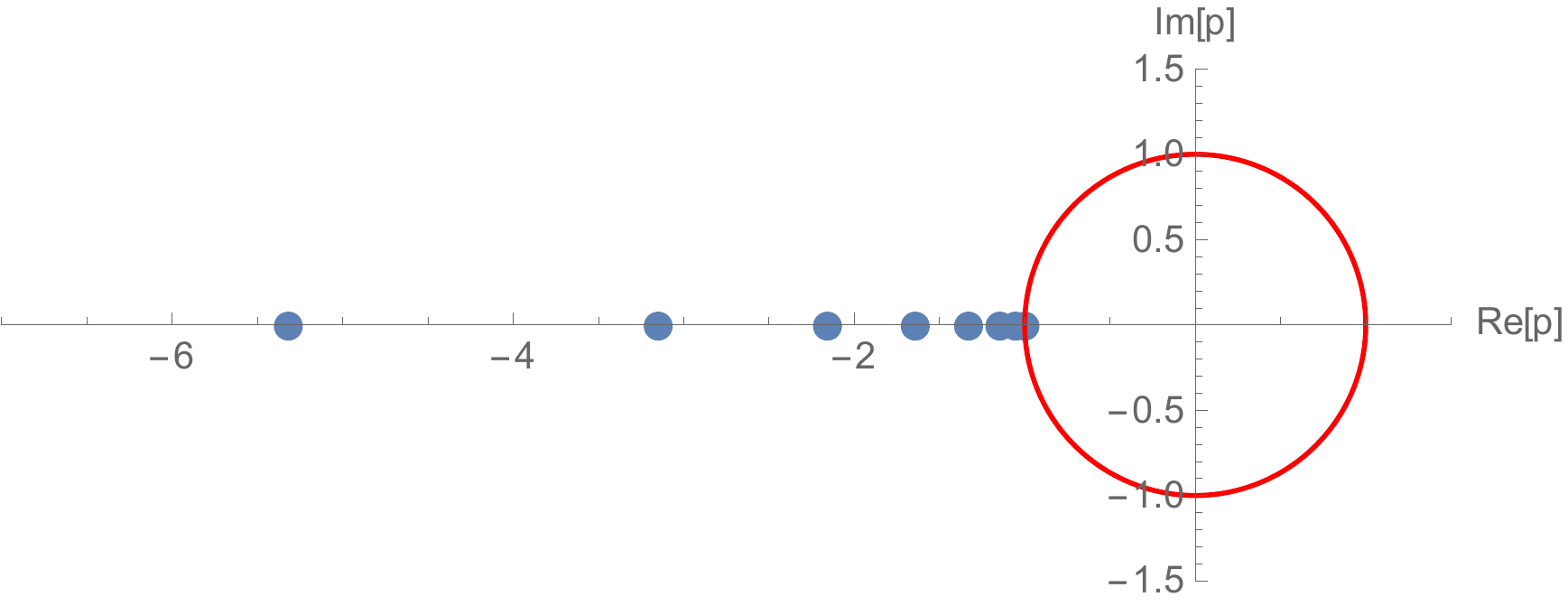}}
\centerline{\includegraphics[scale=0.55]{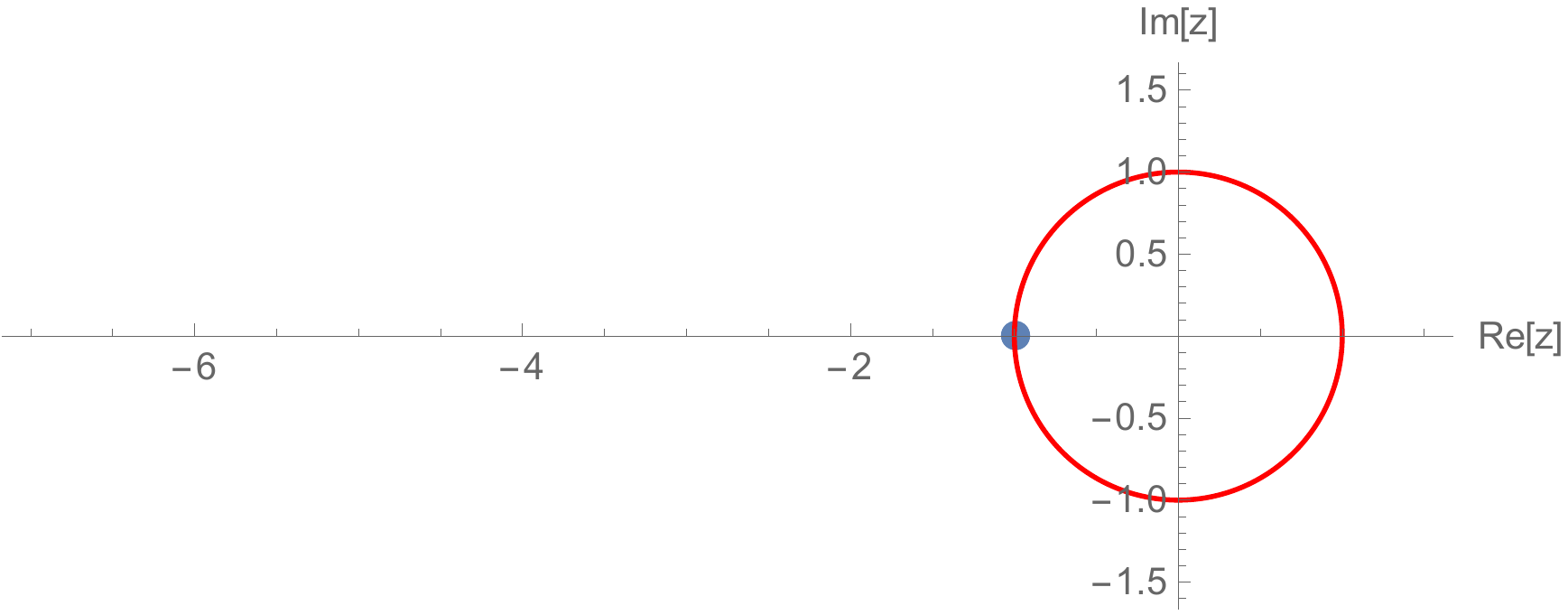}}
\caption{The first plot shows the Pad\'e poles for the $\mathcal{PB}$ approximation to the Borel function $\mathcal{B}(p)=(1+p)^{-1/2}$, from 20 terms of the small $p$ expansion. The square root branch cut is represented approximately as a line of Pad\'e poles accumulating to the branch point at $p=-1$. The second plot shows the Pad\'e poles in the conformally mapped $z$ plane, with the same input, but after the conformal map in (\ref{eq:cmap}). The branch point is mapped to a pole, and Pad\'e is in fact exact.}
\label{fig:square-root}
\end{figure}

\noindent{\bf Comments:} 
\begin{itemize}[label=$\circ$]
\item
We stress that the $\mathcal{PCB}$ result (\ref{eq:pcbp}) is obtained from {\it exactly} the same input information ($2N$ terms of the truncated asymptotic series) as the $\mathcal{PB}$ result (\ref{eq:pbp}), but is significantly more accurate.

\item
This increased precision is particularly dramatic near the Borel cut. See Figure \ref{fig:hyper}. The Pad\'e-Borel transform (\ref{eq:pbp}) places unphysical poles along the negative $p$ axis, $p\in (-\infty, -1]$, accumulating to the branch point at $p=-1$, in an attempt to use rational functions to approximate the natural branch cut of the exact Borel transform function \cite{baker,Stahl,Saff,Yattselev,Costin:2020pcj}. On the other hand, the Pad\'e-Conformal-Borel transform (\ref{eq:pcbp}) has no unphysical poles along the negative $p$ axis, and hence represents the behavior near the Borel branch point and branch cut much more precisely, as is quantified in \cite{Costin:2020hwg} using the large $N$ asymptotics of the  relevant Jacobi polynomials. 

\item
The conformal map (\ref{eq:cmap}) does not require knowledge of the {\it nature} of the branch point, only  its {\it location} (which we have re-scaled here to lie at $p=-1$). Correspondingly, the {\it leading} large $N$ asymptotics of the Pad\'e polynomials is independent of the exponent $\alpha$. However, square root branch points ($\alpha=\pm \frac{1}{2}$) are special, as  in this case the Pad\'e-Conformal-Borel transform is exact for all $N$. The conformal map converts a square root branch point to a pole. See Figure \ref{fig:square-root}. This extreme sensitivity leads to the {\it singularity elimination} method \cite{Costin:2020pcj}, which can be  applied iteratively to obtain remarkably precise knowledge of the exact {\it location} and  {\it exponent} of a Borel singularity.

\item
 In practical computations, it is often more stable numerically  to convert the Pad\'e approximant in the $z$ plane to a partial fraction expansion, and then map back to the Borel $p$ plane. 
 \end{itemize}

{\bf Pad\'e-Uniformized-Borel transform:} if there is a dominant Borel branch point singularity, an even more accurate analytic continuation of the truncated Borel transform function (and hence a more accurate summation of the associated asymptotic series) is obtained by using the uniformizing map of the cut plane:
\begin{eqnarray}
p=-1+e^{s} \quad \longleftrightarrow \quad s=\log(1+p)
\label{eq:uni1}
\end{eqnarray}
This uniformizing map $p=-1+e^{s}$ is a $1$-$1$ map between the complex plane and 
the infinite sheeted Riemann surface of the logarithm function $\log(1+p)$ on ${\mathbb C}$ with the singular point at $p=-1$ removed \cite{abikoff,schlag,Costin:2020pcj}. The point $p=-1$ (the singularity of the logarithm) is a boundary point of this Riemann surface, and is mapped into the boundary point at infinity of the complex $s$ plane (the singularity of the exponential function).

\begin{figure}[htb]
\centerline{\includegraphics[scale=0.6]{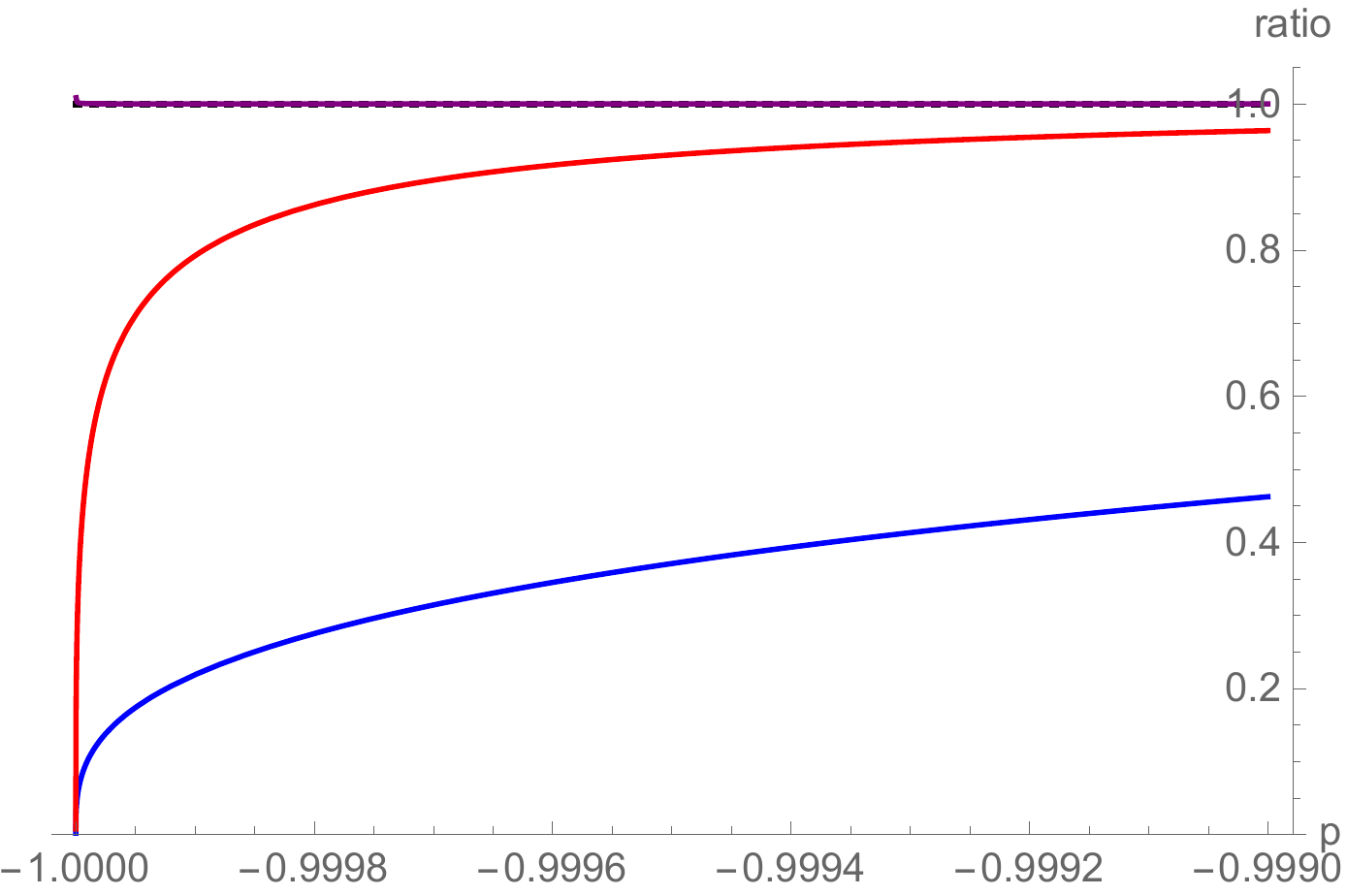}}
\caption{Ratio of $[5,5]$ approximants to the exact Borel transform function $\mathcal{B}(p)= (1+p)^{-1/3}$, as $p\to -1^+$. Blue=Pad\'e-Borel; Red=Pad\'e-Conformal-Borel; Purple= Pad\'e-Uniformizing-Borel. Note the extreme precision of the $\mathcal{PUB}$ approximation compared to the other approximations, especially in the vicinity of the singularity at $p=-1$. Note that each approximation was generated with  the same input data: just 10 terms of the truncated expansion.}
\label{fig:one-cut}
\end{figure}

For the incomplete gamma function example (\ref{eq:i-gamma}), 
the exact diagonal Pad\'e approximant is
\begin{eqnarray}
{\rm Pade}{\rm -}{\rm Uniformized}{\rm -}{\rm Borel:} \quad  {\mathcal {PUB}}_{[N, N]}(p; \alpha)=
\frac{~_1F_1(-N,-2N; \alpha \log(1+p))}{~_1F_1(-N,-2N; -\alpha \log(1+p))}
\label{eq:pubp}
\end{eqnarray}
Once again, large $N$ asymptotics shows that this  Pad\'e-Uniformized-Borel ($\mathcal {PUB}$) transform is more precise than either  the Pad\'e-Borel  or  Pad\'e-Conformal-Borel transform. The improvement in precision is especially dramatic in the vicinity of the Borel singularity: see Figure \ref{fig:one-cut}.
\\

\noindent{\bf Comments:} 
\begin{itemize}[label=$\circ$]

\item
The improvement due to the uniformizing map is particularly dramatic near the singularity. See Figure \ref{fig:one-cut}.

\item
This uniformizing map approach to analytically continuing the truncated Borel transform function is in fact the {\it optimal} extrapolation,  in the sense that this extrapolation constructs the best approximant (i.e., with minimized errors) within the class of all functions analytic on a common Riemann surface (and with common bounds). For details see \cite{Costin:2020pcj}.

\item

Analogous results are straightforwardly obtained for a Borel transform with a logarithmic branch cut, $\mathcal B(p)=\ln(1+p)$, by taking derivatives with respect to the exponent $\alpha$.

\item
While analytic results for the precision were obtained in \cite{Costin:2020hwg} for the asymptotic series having the generic leading large order factorial growth, the incomplete gamma function example in (\ref{eq:i-gamma}), the results apply more generally \cite{Costin:2019xql,Costin:2020pcj}. For example, the hypergeometric Borel transform function in (\ref{eq:f-hyper}) has the same hierarchy of the quality of representations. See for example Figure \ref{fig:hyper}. 

\item
Recall that more precise analytic knowledge of the Borel transform $\mathcal{B}(p)$ yields  more precise analytic knowledge of the physical function $f(x)$ in (\ref{eq:laplace}). In fact,  the improved analytic continuation of $\mathcal{B}(p)$ from the $\mathcal {PUB}$ approximation also permits analytic continuation onto higher Riemann sheets, which is not possible with the other methods \cite{Costin:2020pcj}.

\item
The uniformization method applies in principle to functions with any number of branch points.

 \end{itemize}

\subsection{Two Borel Singularities}
\label{sec:pair}

In applications there often exist two Borel singularities (or two  dominant Borel singularities). In this case there are also explicit conformal maps and uniformizing maps which improve significantly the precision of the analytic continuation of the Borel transform.

\subsubsection{Two Symmetric Collinear Borel Singularities}

\begin{figure}[htb]
\includegraphics[scale=0.45]{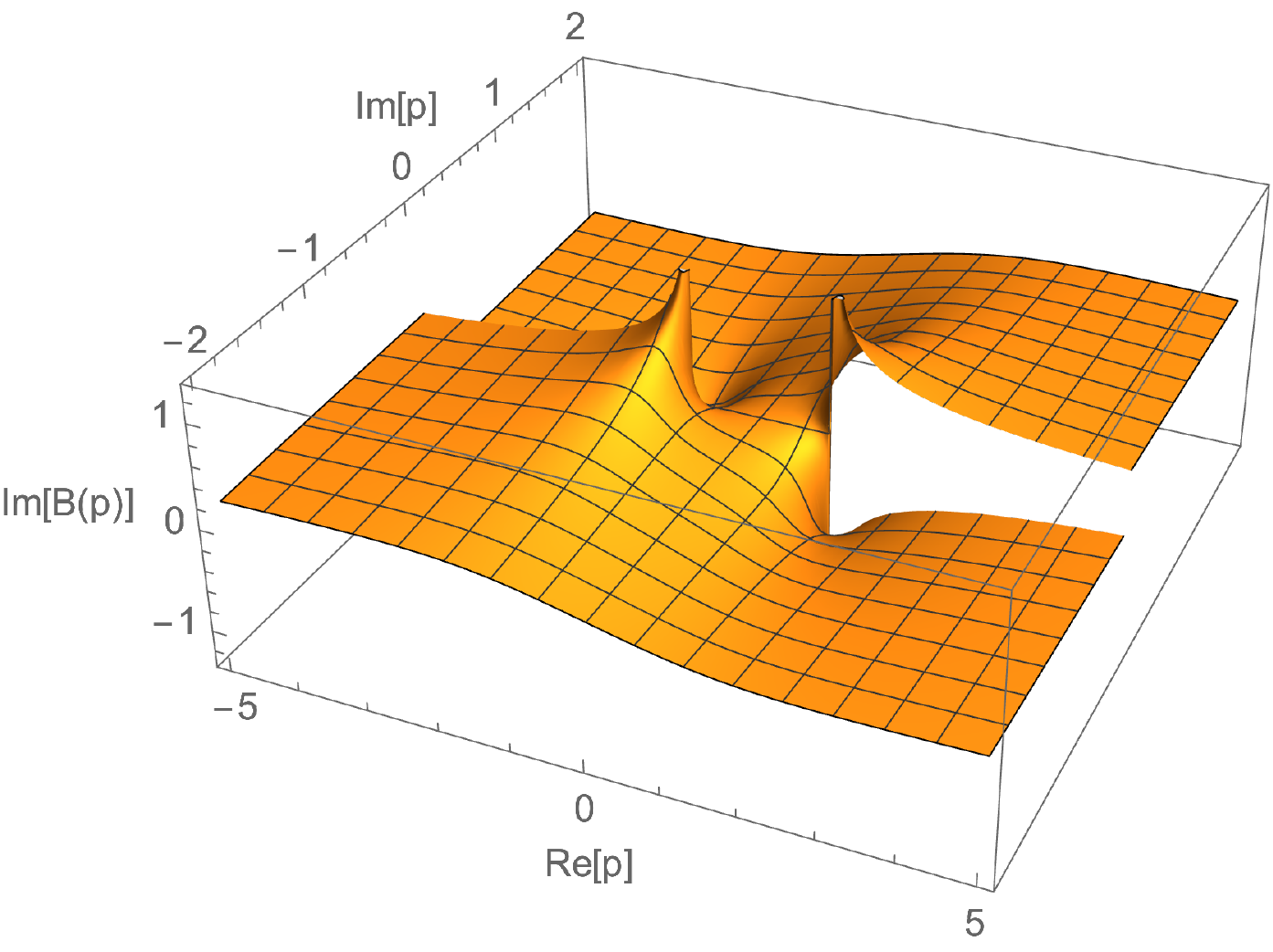}
\includegraphics[scale=0.45]{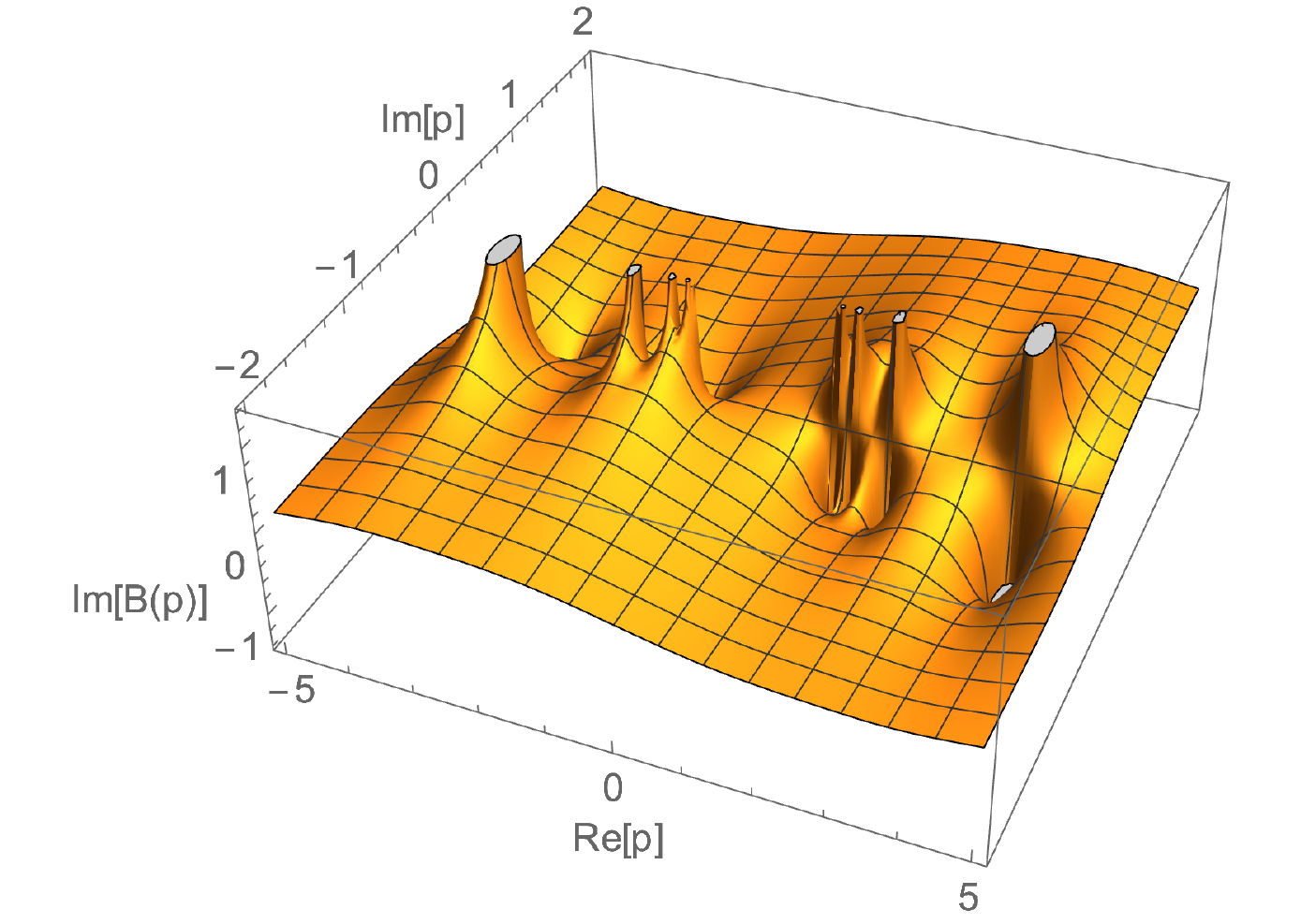}
\includegraphics[scale=0.45]{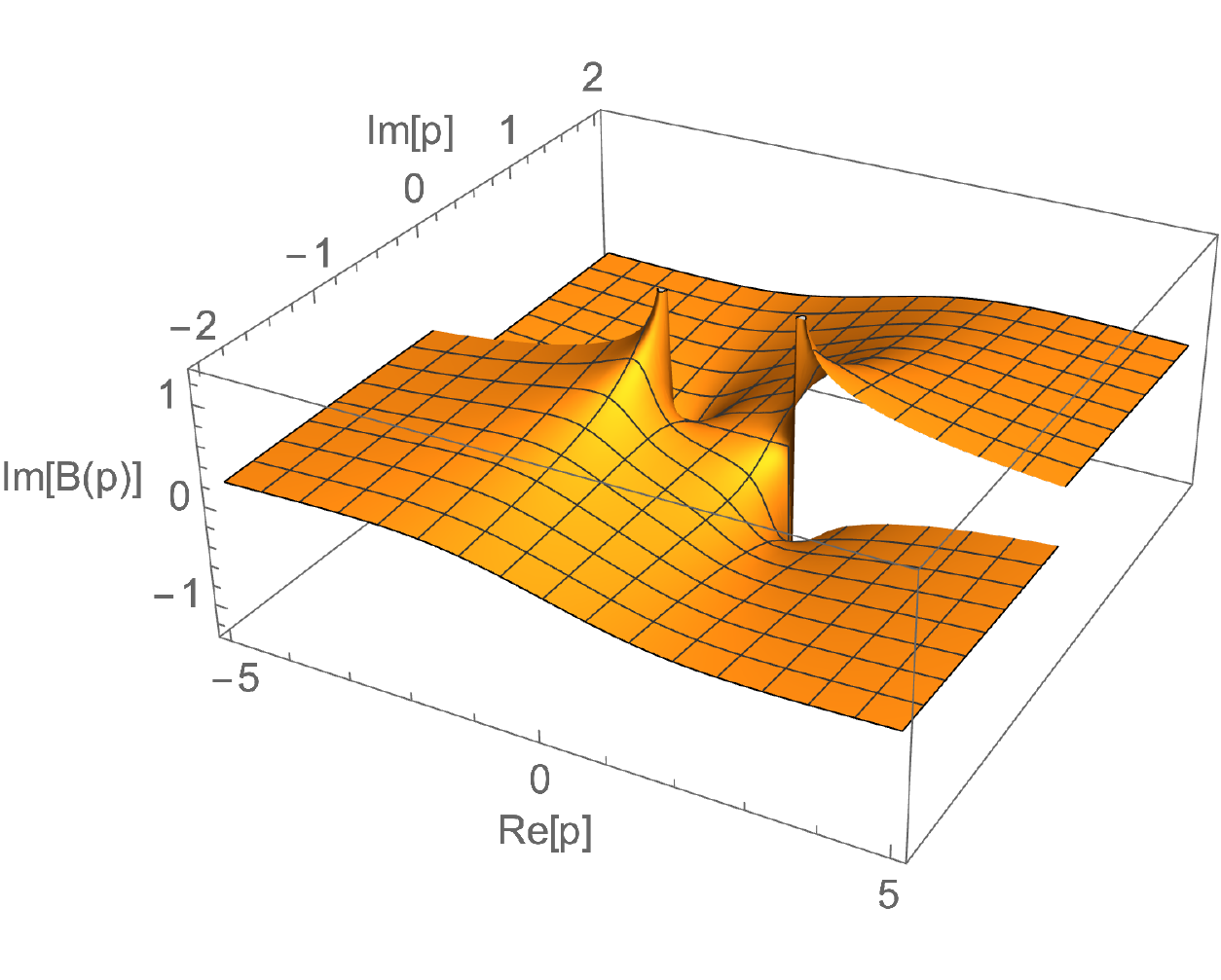}
\includegraphics[scale=0.45]{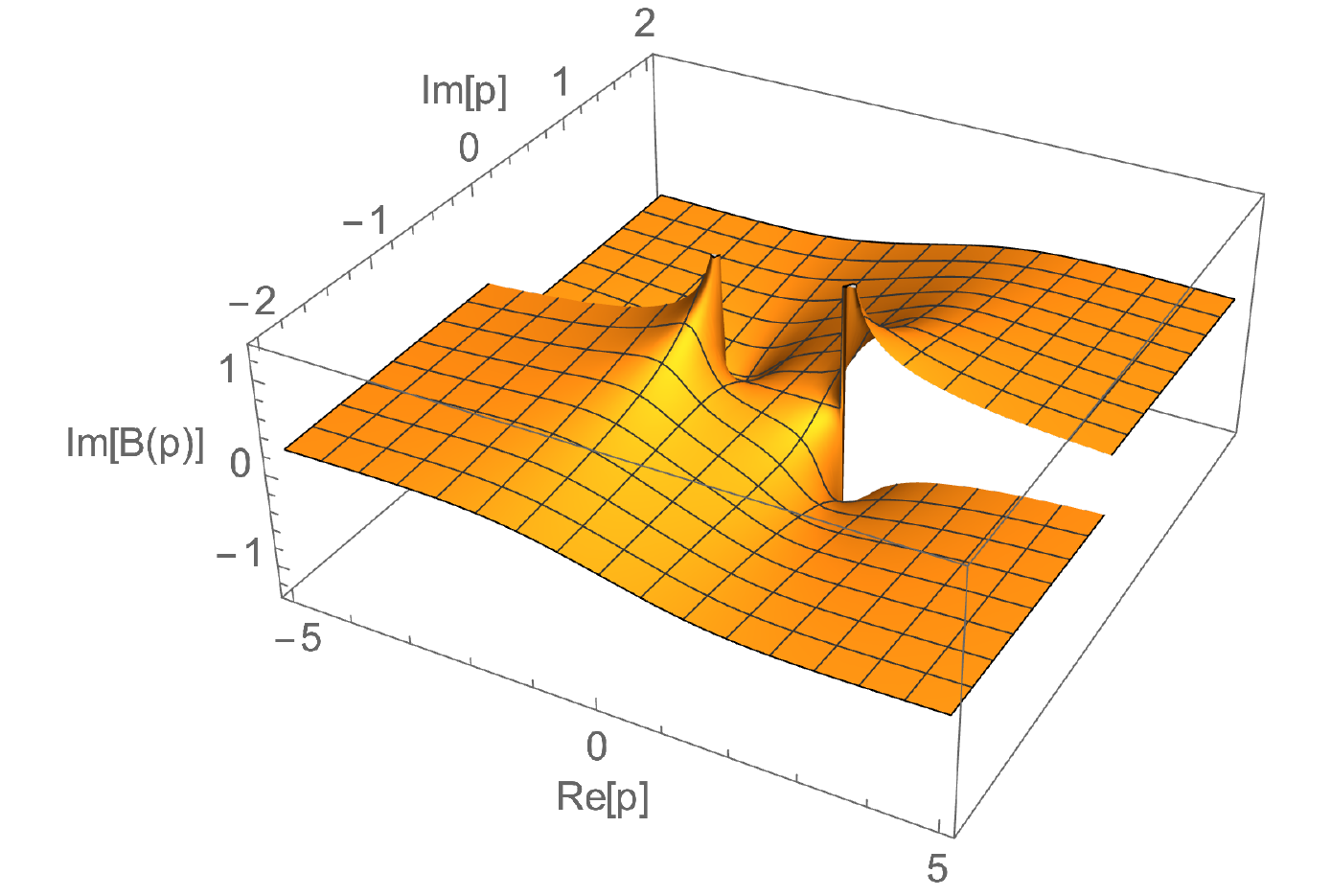}
\caption{Plots of  the imaginary part of the Borel function $\mathcal{B}(p)=(1-p^2)^{-1/3}$ for (i) the exact function; (ii) [10,10] diagonal Pad\'e-Borel approximant;  (iii) [10,10] diagonal Pad\'e-Conformal-Borel  approximant;  (iv) [10,10] diagonal Pad\'e-Uniformized-Borel  approximant. In (ii) Pad\'e places unphysical poles along the branch cuts, while the $\mathcal{PCB}$ and $\mathcal{PUB}$ approximations are much more accurate in the vicinity of the branch cuts.}
\label{fig:2cut}
\end{figure}

Suppose we have two symmetric singularities in the Borel plane, scaled to be at $p=\pm 1$. A simple example is the Borel transform function: 
\begin{eqnarray}
B(p; \alpha)=\left(1-p^2\right)^\alpha
\label{eq:borel2}
\end{eqnarray}
In this case, the conformal map in (\ref{eq:cmap}) generalizes to 
\begin{eqnarray}
p=\frac{2 z}{(1+z^2)}
\quad \longleftrightarrow \quad 
 z=\sqrt{\frac{1-\sqrt{1-p^2}}{1+\sqrt{1-p^2}}}
\label{eq:cmap2}
\end{eqnarray}
The $\mathcal{PB}$ and $\mathcal{PCB}$ approximations generalize in a straightforward way [compare with
(\ref{eq:pbp}) and (\ref{eq:pcbp})]:
\begin{eqnarray}
{\rm Pade}{\rm -}{\rm Borel:} \qquad {\mathcal {PB}}_{[N,N]}(p; \alpha)=\frac{P_N^{(\alpha,- \alpha)}\left(1-\frac{2}{p^2}\right)}{P_N^{(-\alpha, \alpha)}\left(1-\frac{2}{p^2}\right)}
\label{eq:pbp2}
\end{eqnarray}
\begin{eqnarray}
{\rm Pade}{\rm -}{\rm Conformal}{\rm -}{\rm Borel:} \qquad {\mathcal {PCB}}_{[N,N]}(p; \alpha)=\frac{P_N^{(2\alpha,- 2\alpha)}\left(\frac{\sqrt{1-p^2}+1}{\sqrt{1-p^2}-1}\right)}{P_N^{(-2\alpha, 2\alpha)}\left(\frac{\sqrt{1-p^2}+1}{\sqrt{1-p^2}-1}\right)}
\label{eq:pcbp2}
\end{eqnarray}
The $\mathcal{PUB}$ approximation uses the uniformizing map for the universal covering of  $\hat{\mathbb C}\setminus \{-1,1, \infty\}$:
\begin{eqnarray}
p =-1+2 \lambda \left(i\, \frac{1-z}{1+z}\right)
\quad \longleftrightarrow \quad
z= \frac{{\mathbb K}\left(\frac{1+p}{2}\right)-{\mathbb K}\left(\frac{1-p}{2}\right)}{{\mathbb K}\left(\frac{1-p}{2}\right)+{\mathbb K}\left(\frac{1+p}{2}\right)} 
\label{eq:uni2}
\end{eqnarray}
Here $\lambda=\theta_2^4/\theta_3^4$ is the elliptic modular function, $\theta_2,\theta_3$ are Jacobi theta functions, and ${\mathbb K}(m)=(\pi/2) \,_2F_1(\frac{1}{2}, \frac{1}{2};1;m)$ is the complete elliptic integral of the first kind of modulus $m=k^2$  \cite{bateman}. 
As before, the $\mathcal{PUB}$ approximation involves mapping from the Borel $p$ plane to the $z$ plane, making a Pad\'e approximant in $z$, and then mapping back to the Borel $p$ plane using the inverse map. The $\mathcal{PB}$ approximation places unphysical poles along the two cuts, while the $\mathcal{PCB}$ and $\mathcal{PUB}$ approximation provide a very accurate representation of the Borel cuts, even using just 20 terms of the original asymptotic expansion. See Figure \ref{fig:2cut}.
\\

\noindent{\bf Comment:} 
\begin{itemize}[label=$\circ$]
\item
The improvement in accuracy with the $\mathcal{PUB}$ approximation is particularly dramatic near the singular points. Indeed, the leading order asymptotic behavior of $z$ near $p=1$ is $z\sim 1+2\pi/\ln(1-p)$. Therefore,
in the vicinity of one of the singularities there is exponential ``stretching'', $p\sim 1-\exp\left[-\frac{2\pi}{1-z}\right]$, which means for example that $z\approx 0.9$ corresponds to $p\approx 1-5\cdot 10^{-28}$. This enables ultra-precise probing of the $p$ plane singularities.

\end{itemize}

\subsubsection{Two Asymmetric Collinear Borel Singularities}

Another common physical configuration of Borel singularities consists of an asymmetric collinear pair of Borel singularities, for example at $p=-a$ and $p=+b$,  with $a$ and $b$ both real,  with natural cuts $p\in [b, \infty)$ and $p\in (-\infty, -a]$ on either side of the real axis. 
The corresponding conformal map is:
\begin{eqnarray}
p=\frac{4\, a\, b\, z}{a (1 + z)^2 + b (1 - z)^2}
\quad \longleftrightarrow \quad
z=\frac{1-\sqrt{\frac{a (b-p)}{b (a+p)}}}{1+\sqrt{\frac{a (b-p)}{b (a+p)}}}
\label{eq:cmap-asymmetric}
\end{eqnarray}

\subsubsection{Complex Conjugate Pair of Borel Singularities}

An important physically relevant configuration of two Borel singularities is a complex conjugate pair, which occurs for example in problems with symmetry breaking \cite{heller,serone,bertrand,rossi}. We can normalize these to lie at $p=e^{\pm i\theta}$, and choose $\theta\in \left[0, \frac{\pi}{2}\right]$.
\\

\begin{figure}[htb]
\centerline{\includegraphics[scale=0.7]{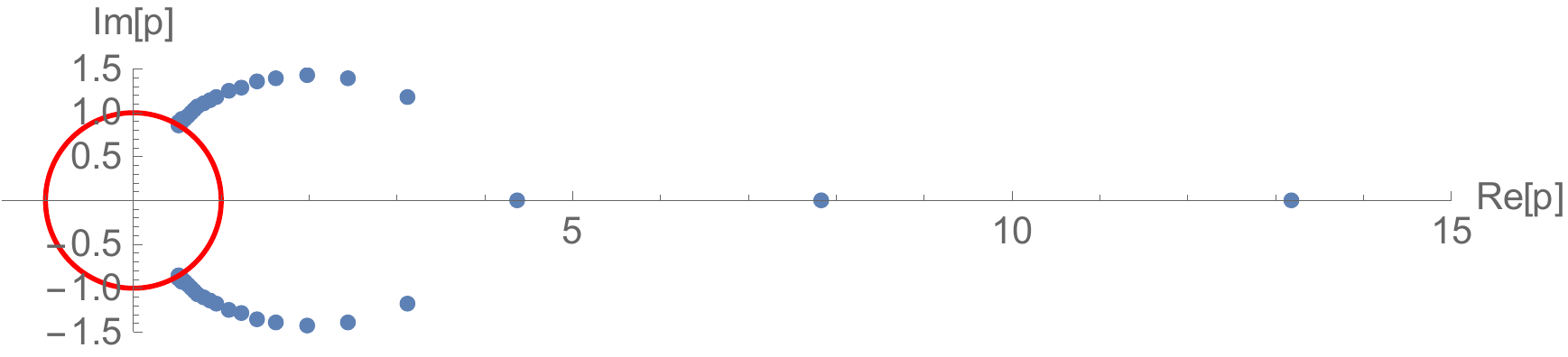}}
\caption{The poles of the Pad\'e approximation to the Borel transform function $\mathcal B(p)=(1-2\cos\left(\frac{\pi}{3}\right)p+p^2)^{-1/3}$, which has a complex conjugate pair of Borel singularities at $p=e^{\pm i \pi/3}$. The unit circle is shown in red. Pad\'e produces arcs of poles emanating from $p=e^{\pm i \pi/3}$, joining and continuing along the positive real $p$ axis. These Pad\'e poles are not related to singularities of the original function $\mathcal B(p)$, and limit the precision of the Borel-Laplace integral in (\ref{eq:laplace}). }
\label{fig:pb-pair}
\end{figure}
\begin{figure}[htb]
\centerline{\includegraphics[scale=0.7]{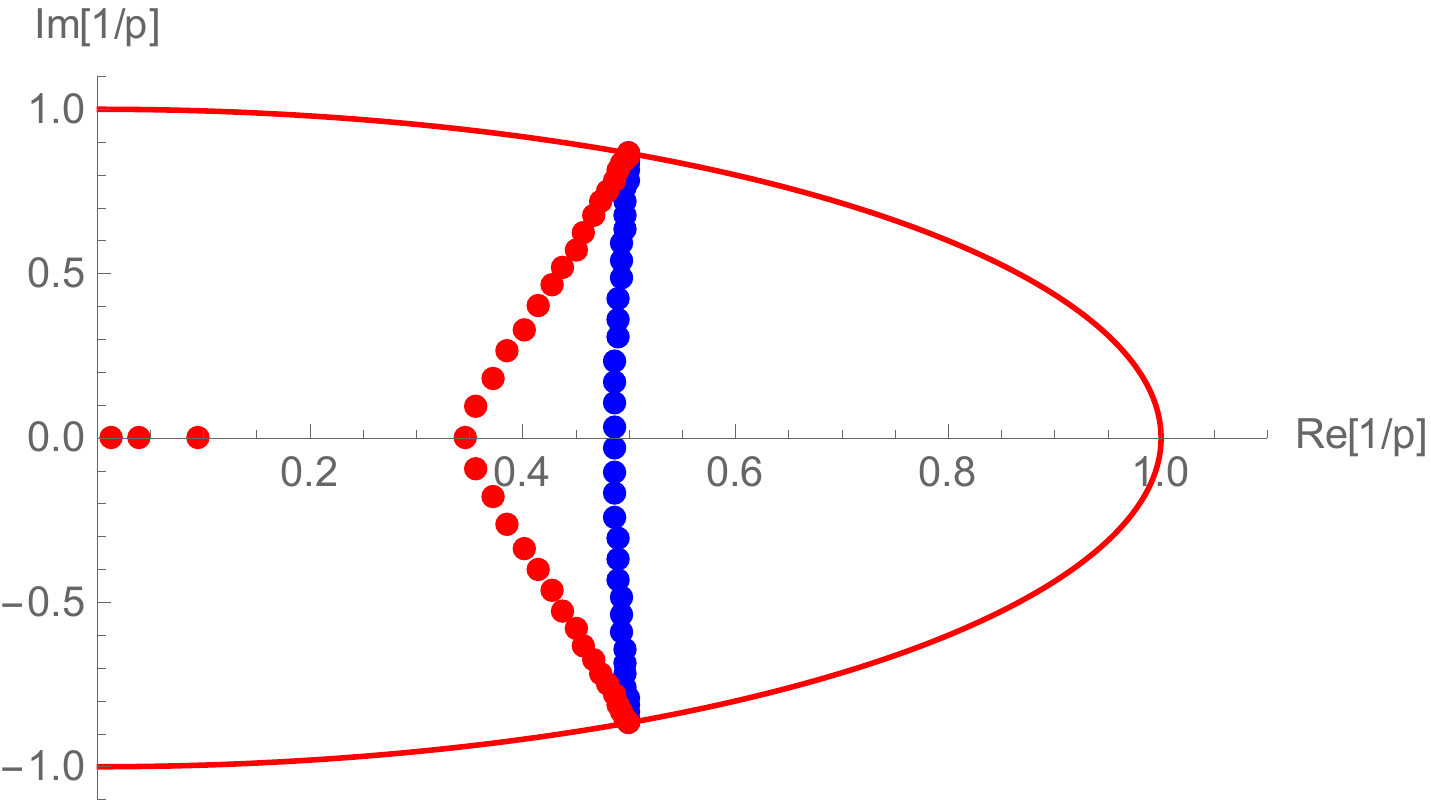}}
\caption{The inverse of the Pad\'e poles for the function $\mathcal B(p)=(1-2\cos\left(\frac{\pi}{3}\right)p+p^2)^{-1/2}$ in blue. These inverse poles show the form of the ``minimal capacitor'' discussed in Section \ref{sec:pade}.  In the limit of infinite Pad\'e order these form a vertical straight line between the two singularities at $p=e^{\pm i \pi/3}$. If the exponent is shifted by a tiny amount, here just $10^{-4}$, the inverse Pad\'e poles (red dots) form a trivalent graph connecting the origin and the two singularities. 
 }
\label{fig:square-root-capacitor}
\end{figure}
\noindent{\bf Comments:} 
\begin{itemize}[label=$\circ$]

\item
The $\mathcal{PB}$ approximation produces two curved arcs of poles, joining to a line of poles along the positive $p$ axis. This illustrates the potential theory interpretation of Pad\'e described in Section \ref{sec:potential}. These unphysical poles limit the precision of the Borel-Laplace integral (\ref{eq:laplace}). See Figure \ref{fig:pb-pair}. 

\item
Pad\'e is extremely sensitive to a square root branch point. See Figure \ref{fig:square-root-capacitor}, which shows that a tiny shift in the exponent away from a square root ($\frac{1}{2}\to \frac{1}{2}-10^{-4}$) has a dramatic and easily recognizable effect on the Pad\'e pole distribution.

\end{itemize}

For this configuration of two Borel singularities at $p=e^{\pm i \theta}$, the conformal map is
\begin{eqnarray}
p= c(\theta)\frac{z}{(1+z)^{2} } \left(\frac{1+z}{1-z}\right)^{2\theta/\pi}
\quad, \quad c(\theta)=4\left(\frac{\theta}{\pi}\right)^{\theta/\pi}\left(1-\frac{\theta}{\pi}\right)^{1-\theta/\pi} 
\label{eq:cmap-pair}
\end{eqnarray}

\begin{figure}[htb]
\centerline{\includegraphics[scale=0.6]{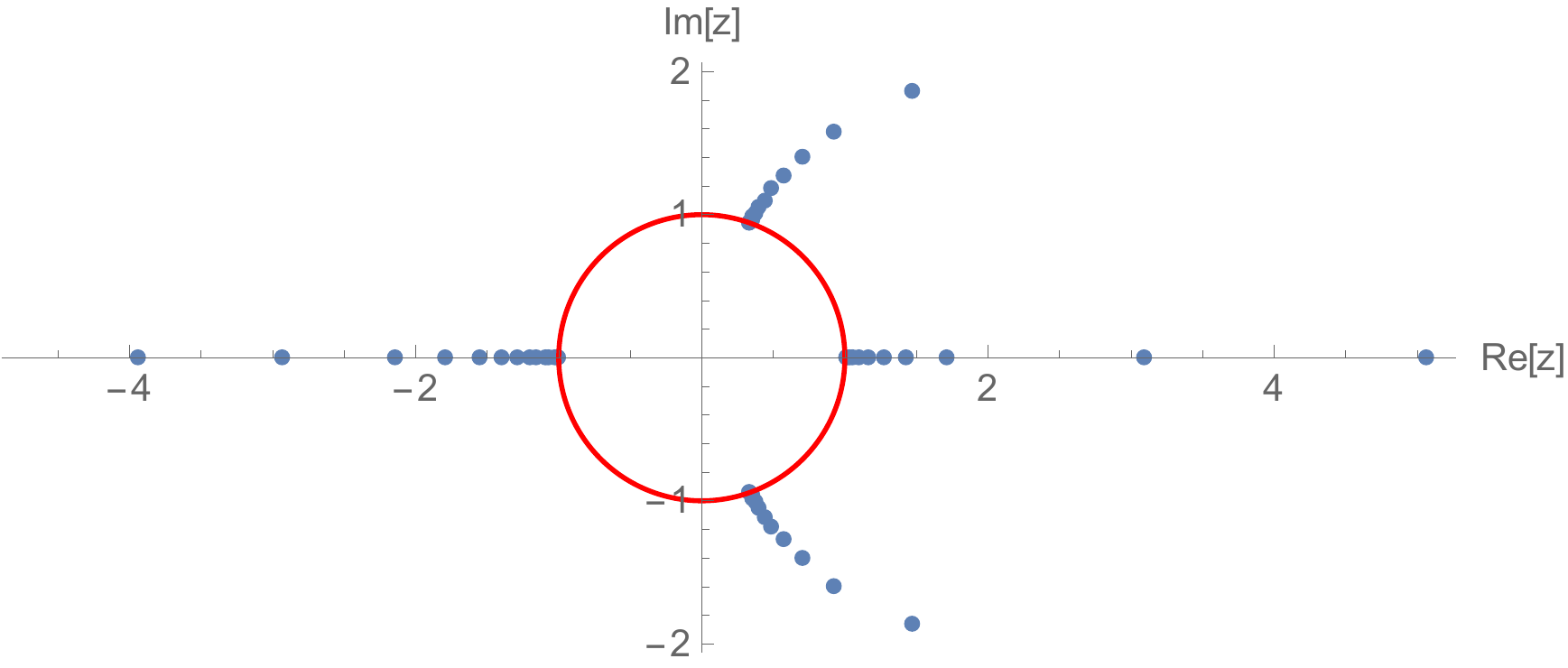}}
\caption{The $z$ plane poles of the Pad\'e approximation to the Borel transform function $\mathcal B(p)=(1-2\cos\left(\frac{\pi}{3}\right)p+p^2)^{-1/3}$, after conformal mapping to the $z$ plane. The unit circle is shown in red. Pad\'e produces arcs of poles emanating from $z=\frac{1}{3}(1\pm 2\sqrt{2} i)$, the conformal map images of $p=e^{\pm i \pi/3}$. The poles accumulating to $z=\pm 1$ correspond to the singularity at $p=\infty$. }
\label{fig:pcbz-pair}
\end{figure}

\noindent{\bf Comments:} 
\begin{itemize}[label=$\circ$]

\item
With the conformal map (\ref{eq:cmap-pair}), the $\mathcal{PCB}$ approximation leads to two symmetric arcs of poles in the conformal $z$ plane, emanating from the conformal map images of $p=e^{\pm i\theta}$. See Figure  \ref{fig:pcbz-pair}. Note that there is now no obstacle to computing the Borel-Laplace integral in the $z$ plane, integrating from $z=0$ to $z=1$.

\item
The conformal map (\ref{eq:cmap-pair}) is explicit, but its inverse is not, except for special simple rational values of $\theta/\pi$.

\end{itemize}

For this configuration of two Borel singularities at $p=e^{\pm i \theta}$, the uniformizing map is
\begin{eqnarray}
p=  e^{i\, \theta} -2i\,\sin(\theta)\,  \lambda\left(i\left(
\frac{{\mathbb K}\left(\frac{1}{2}+\frac{i}{2} \cot\theta\right)-{\mathbb K}\left(\frac{1}{2}-\frac{i}{2} \cot\theta\right) z}
{{\mathbb K}\left(\frac{1}{2}-\frac{i}{2} \cot\theta\right)+{\mathbb K}\left(\frac{1}{2}+\frac{i}{2} \cot\theta\right) z}\right)\right)
\label{eq:umap-pair}
\end{eqnarray}
where $\lambda$ is the modular $\lambda$ function \cite{bateman}. The inverse map is explicit:
\begin{eqnarray}
z=\frac{Z(p; \theta) -Z(0; \theta)}{1-\left(Z(0; \theta)\right)^*\, Z(p; \theta)}
\label{eq:cc2}
\end{eqnarray}
where $Z(p; \theta)$ is defined as
\begin{eqnarray}
Z(p; \theta)\equiv \frac{{\mathbb K}\left(\frac{1}{2}+\frac{i}{2}\left(\frac{p}{\sin\theta}-\cot \theta\right)\right)
-{\mathbb K}\left(\frac{1}{2}-\frac{i}{2}\left(\frac{p}{\sin\theta}-\cot \theta\right)\right)}{{\mathbb K}\left(\frac{1}{2}+\frac{i}{2}\left(\frac{p}{\sin\theta}-\cot \theta\right)\right)
+{\mathbb K}\left(\frac{1}{2}-\frac{i}{2}\left(\frac{p}{\sin\theta}-\cot \theta\right)\right)}
\label{eq:cc2Z}
\end{eqnarray}

\noindent{\bf Comments:} 
\begin{itemize}[label=$\circ$]
\item
This uniformizing map and its inverse are both explicit, and it is also significantly more precise than the conformal map (\ref{eq:cmap-pair}), especially near the singularities.

\item
It is straightforward to generalize the uniformizing map of the previous examples to the case of two non-symmetric complex Borel singularities, $p_1, p_2 \in {\mathbb C}$, using a suitable M\"obius transformation and disk automorphism to obtain the universal covering of $\hat{\mathbb C}\setminus \{p_1, p_2, \infty\}$. The uniformization maps are again expressed in terms of the elliptic function ${\mathbb K}$ and the elliptic modular function $\lambda$. 

\end{itemize}

\subsection{Three or More Borel Singularities}
\label{sec:multi}

\subsubsection{$k$-fold Symmetrically Distributed  Borel Singularities}

\begin{figure}[htb!]
\centerline{\includegraphics[scale=0.45]{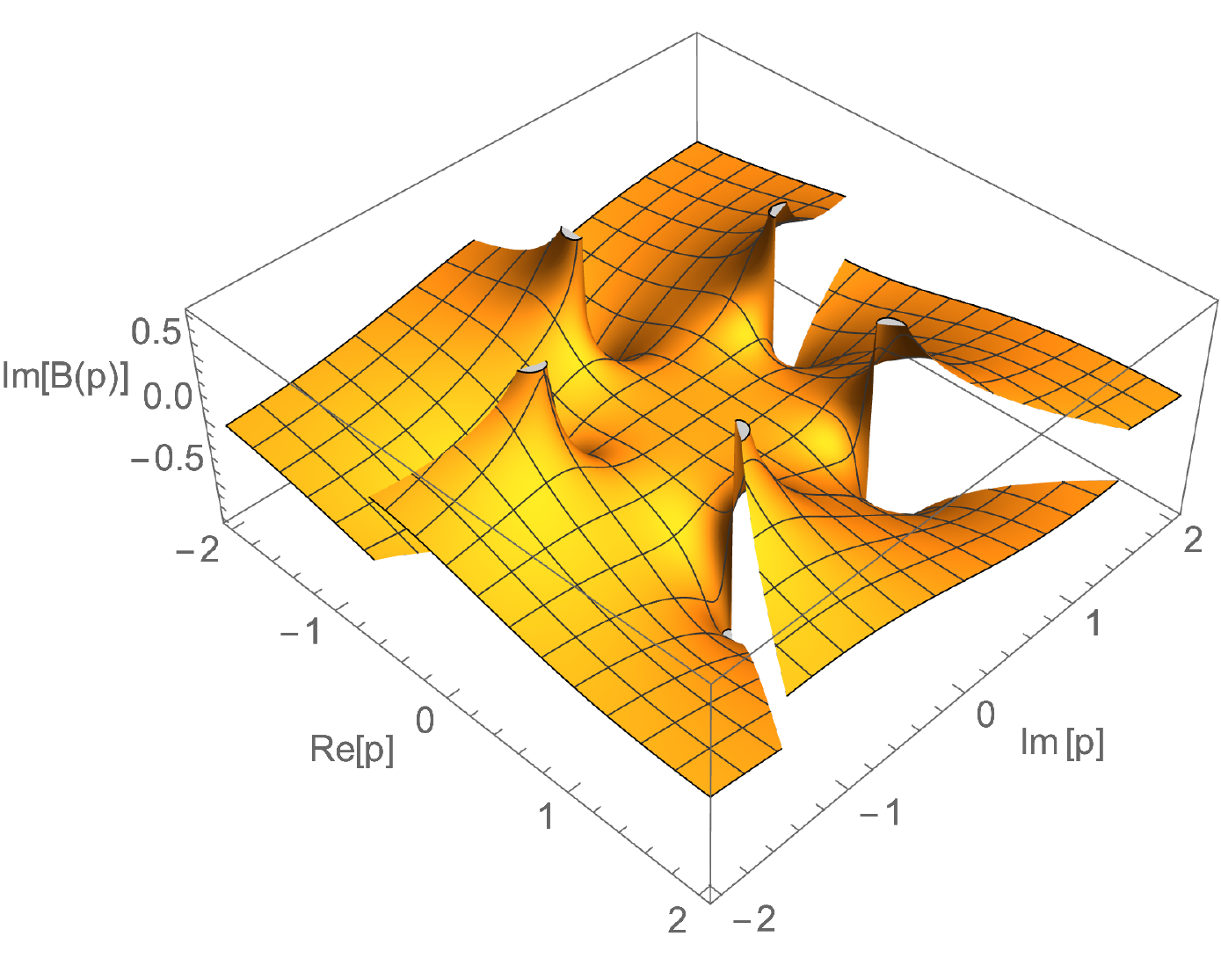}
\quad 
\includegraphics[scale=0.45]{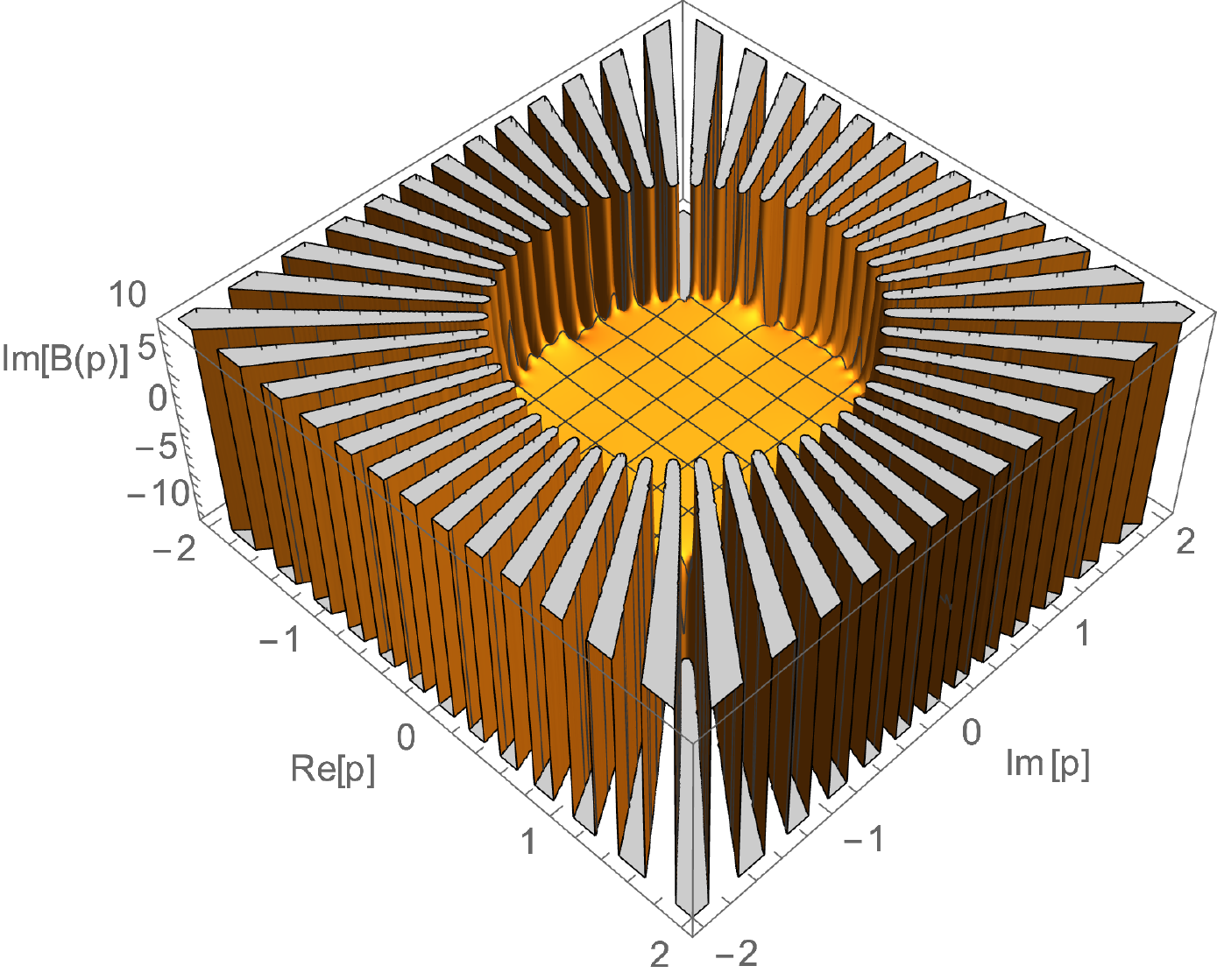}
}
\centerline{\includegraphics[scale=0.45]{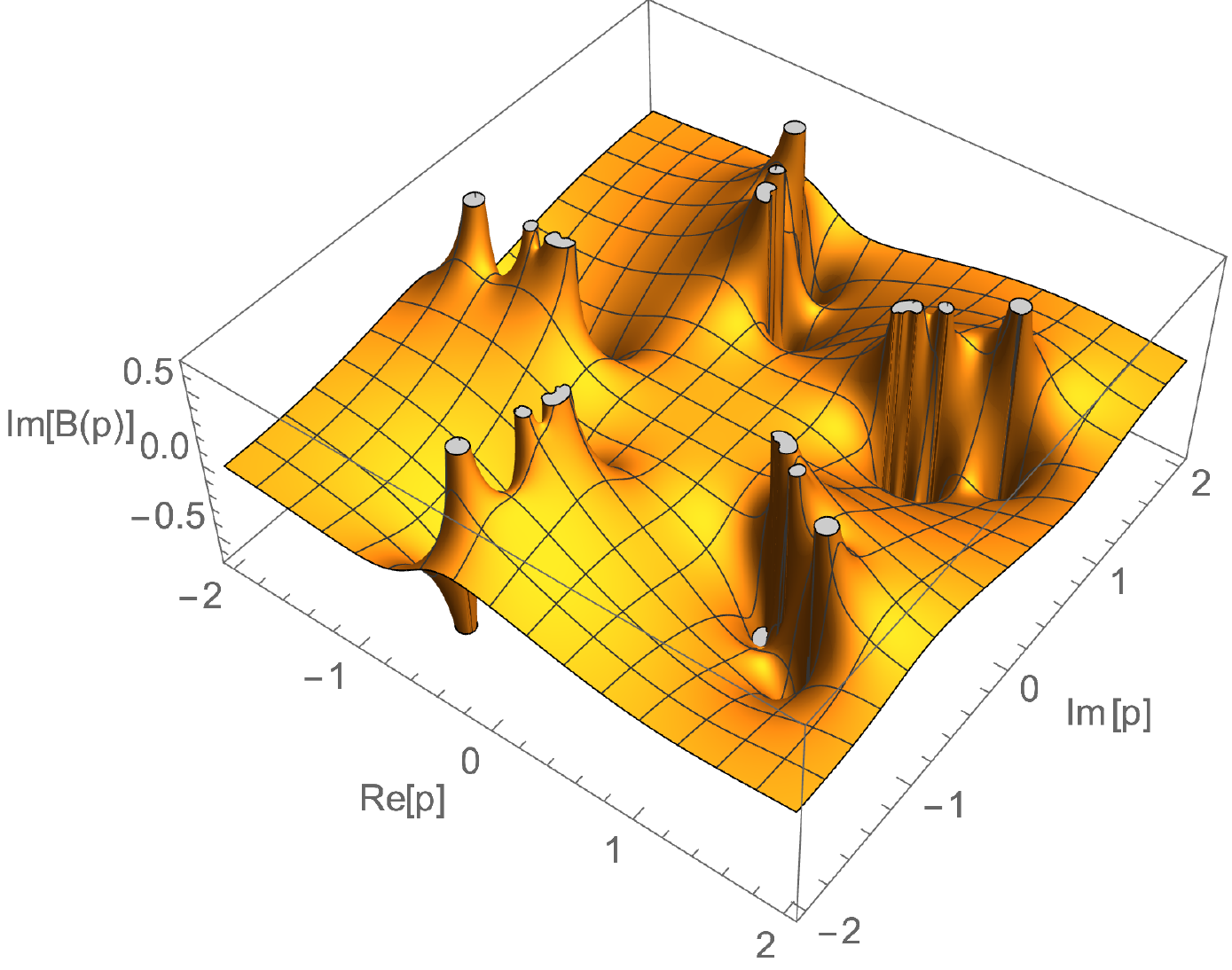}
\quad 
\includegraphics[scale=0.45]{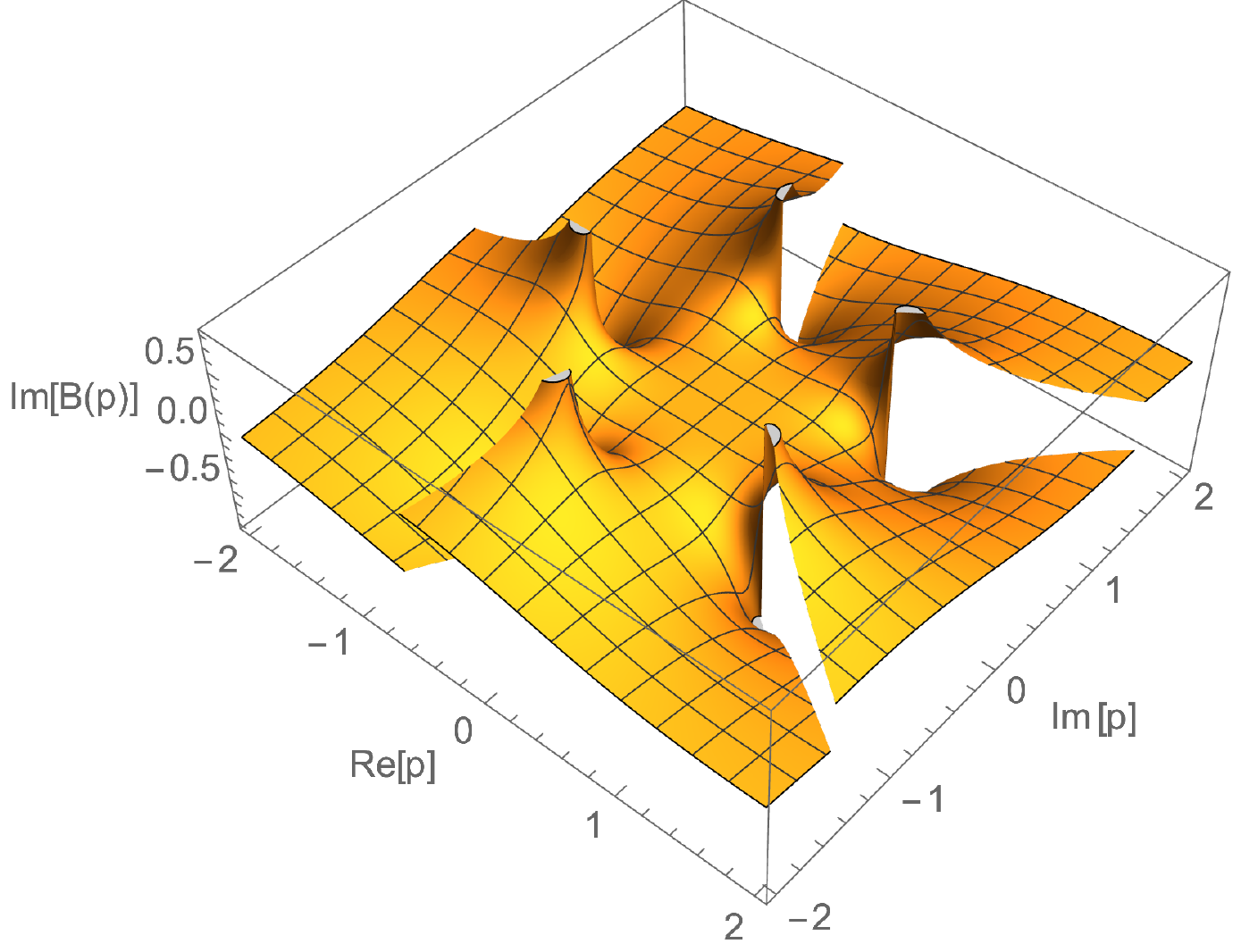}
}
\caption{Plots of the imaginary part of the Borel function $\mathcal{B}(p)=(1+p^5)^{-1/3}$, which has 5 symmetric branch points at $p=e^{(2n+1) i \pi/5}$, for $n=0, 1, 2, 3, 4$. The first plot is for the exact function, showing five symmetric radial cuts; the second plot is for 10 terms of the truncated series; the third plot is the Pad\'e-Borel approximation of the 10-term truncated series; the fourth plot is the Pad\'e-Conformal-Borel approximation of the 10-term truncated series. The truncated series is useless beyond $|p|<1$. The $\mathcal{PB}$ approximation extrapolates well away from the cuts, but places unphysical poles along the cuts, while the $\mathcal{PCB}$ approximation is very accurate also near the cuts, starting with just 10 terms.}
\label{fig:5cut}
\end{figure}

The general problem of constructing conformal and uniformizing maps with more singularities is a non-trivial problem, even numerically \cite{bateman,Nehari,hempel,kober,trefethen}.  However, in physical applications the Borel singularities are often distributed symmetrically, in which case more can be done.
For example, consider a symmetric $k$-fold set of singularities emanating from the vertices of a regular polygon, such as for the Borel transform function:
\begin{eqnarray}
\mathcal{B}(p; \alpha, k)=\left(1+p^k\right)^\alpha
\label{eq:bk}
\end{eqnarray}
The conformal map (\ref{eq:cmap}) 
generalizes to 
\begin{eqnarray}
p=\frac{2^{2/k} z}{(1-z^k)^{2/k}}
\quad \longleftrightarrow \quad 
z^k=\frac{\sqrt{1+p^k}-1}{\sqrt{1+p^k}+1}
\label{eq:cmap-k}
\end{eqnarray}
with natural branch choices.
The $\mathcal{PB}$ and $\mathcal{PCB}$ approximations generalize in a straightforward way, with $p$ simply replaced by $p^k$ in the expressions (\ref{eq:pbp}) and (\ref{eq:pcbp}). See Figure \ref{fig:5cut}.
\\

\noindent{\bf Comments:} 
\begin{itemize}[label=$\circ$]
\item
There is a more general principle underlying this construction. Suppose $p(z)$ is a conformal map of  a cut region $\mathcal D\in \hat{\mathbb C}$ into the unit disk, with $p'(z=0)>0$. Then the map $p_k(z)=\left(p(z^k)\right)^{1/k}$ is the conformal map for $k$ symmetric copies of $\mathcal D^{1/k}$ \cite{Costin:2020pcj}.

\item

The same result applies to uniformizing maps.

\end{itemize}

\subsubsection{Schwarz-Christoffel Construction}

For more general configurations of singularities,  Schwarz-Christoffel provides a constructive approach \cite{Nehari}.
For this discussion it is simpler to invert ($p\to 1/p$) to move the point of analyticity from $p=0$ to $p=\infty$. 
This choice is motivated physically by the electrostatic potential interpretation of Pad\'e approximants  \cite{Stahl,Saff,Costin:2020pcj}, in which the potential at infinity is taken to vanish.
Then for a general \underline{finite} set of branch points, $S=\{p_1,...,p_n\}$, Pad\'e produces a conformal map (in the infinite Pad\'e order limit) which corresponds to the minimal capacitor \cite{Stahl,Saff,Costin:2020pcj}. The expansion at infinity can be written
\begin{equation}
  \label{eq:confomapinfi}
\mathcal B(p)=C_B p+\sum_{k=0}^\infty b_k p^{-k}
\end{equation}
where $C_B$ is the capacity. Using potential theory, there exists a set $\{a_1,...,a_{n-2}\}$ of (complex) auxiliary parameters
such that $\mathcal B(p)$ satisfies
\begin{equation}
  \label{eq:confomapeq}
 \log p=\int^{\mathcal B(p)}\sqrt{\frac{\prod_{j=1}^{n-2}(s-a_j)}{\prod_{j=1}^{n}(s-p_j)}}ds
\end{equation}
Geometrically, the auxiliary parameters are the intersection points of the set of analytic arcs of 
 poles of the diagonal Pad\'e approximation 
 {\em for any} function $F$ having $S$ as the set of branch points, and being analytic in the complement of the minimal capacitor. 
See for example the red dots in Figure \ref{fig:probe}, which have accumulation points at $p=-1$, $p=-2$ and $p=-\frac{3}{2}+\frac{i}{2}$, corresponding to the singularities at  these points, and also have two trivalent vertices near $p=-2+\frac{3i}{20}$ and $p=-\frac{3}{2}+\frac{3i}{10}$. The inverses of these trivalent vertices are the the auxiliary parameters in this case.
The analytic arcs $\gamma_j$ ($\gamma'_j$, resp)  joining $a_1$ with $p_j, j=1,...,n-1$ ($a_1$ to $a_j, j=2,...,n-2$, resp.) are given by the reality conditions:
\begin{equation}
  \label{eq:arcs}
{\rm Re}\left[\int_{\gamma_k}^{p}\sqrt{\frac{\prod_{j=1}^{n-2}(s-a_j)}{\prod_{j=1}^{n}(s-p_j)}}ds\right]=0
\quad \& \quad  
{\rm Re}\left[\int_{\gamma'_m}^{p}\sqrt{\frac{\prod_{j=1}^{n-2}(s-a_j)}{\prod_{j=1}^{n}(s-p_j)}}ds\right]=0
\end{equation}
where $k=1,...,n-1$ and $m=2,...,n-2$. In cases of symmetrically distributed branch points, these integrals can be expressed in terms of elementary or elliptic functions 
 \cite{kuzmina}, and in more general cases the minimal capacitor produced by Pad\'e can be found numerically
\cite{grassmann,trefethen,Crowdy}.

\subsubsection{Repeated Borel Singularities}
\label{sec:repeated}

For nonlinear problems, such as nonlinear ODEs and PDEs, a given Borel singularity is typically repeated an infinite number of times, often in integer multiples along a ray from the origin. For example, the Borel plane singularities of solutions to nonlinear ODEs, such as the Painlev\'e equations, lie in integer multiples along the real Borel axis, and this can be uniformized using elliptic functions \cite{Costin:2020pcj}.\footnote{These maps can also be approximated by rapidly convergent iterations of simple maps \cite{Costin:2020pcj}.} 
 In other cases the Borel singularities are repeated in $({\rm integer})^2$ multiples \cite{Gukov:2016njj}, or in parabolic arrays \cite{mckean}.
  \begin{figure}[htb]
\centerline{\includegraphics[scale=0.75]{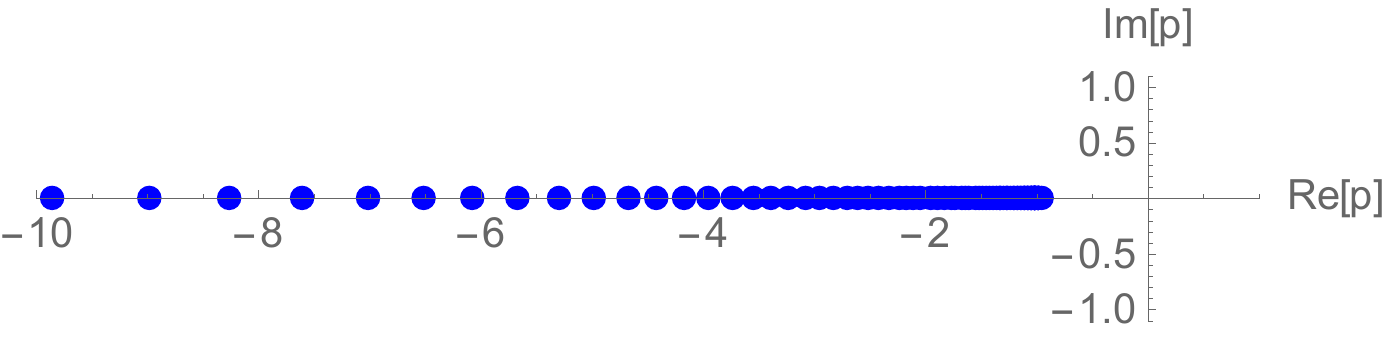}}
\centerline{\includegraphics[scale=0.7]{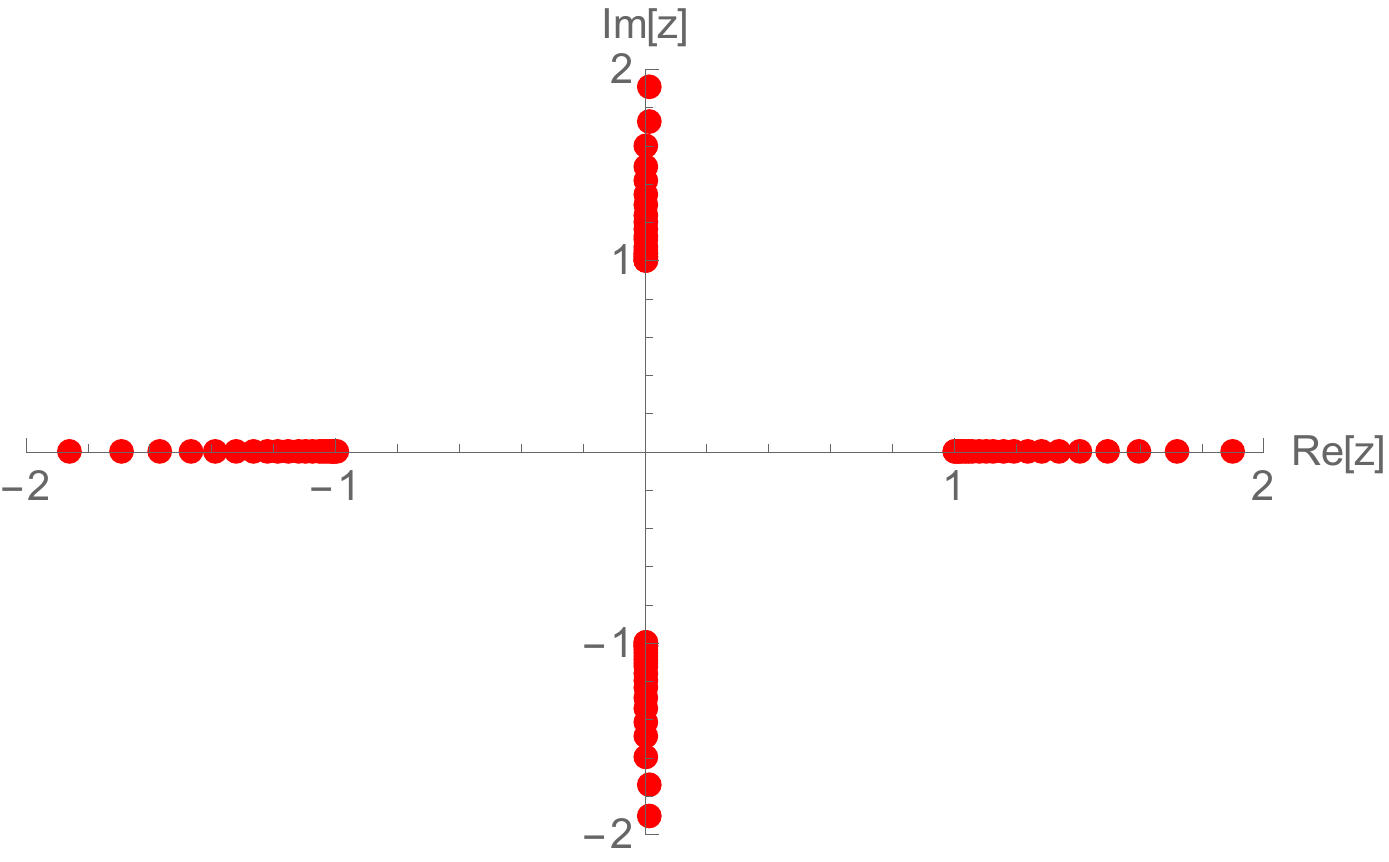}}
\caption{The upper figure shows the Pad\'e poles (blue) of the  Borel function $B(p)$ in (\ref{eq:probe1}), indicating a branch point at $p=-1$. The singularity at $p=-2$ is obscured by the Pad\'e poles representing the branch cut $p\in (-\infty,-1]$. After the conformal map (\ref{eq:cmap}) based on the leading singularity at $p=-1$, the Pad\'e poles in the $z$ plane (red) indicate singularities at $z=\pm 1$ (which are the conformal map images of $p=-1$ and $p=\infty$), and also at $z=\pm i$ (which are the images of $p=-2$ on both sides of the cut).}
\label{fig:hidden}
\end{figure}
\begin{figure}[htb]
\centerline{\includegraphics[scale=0.7]{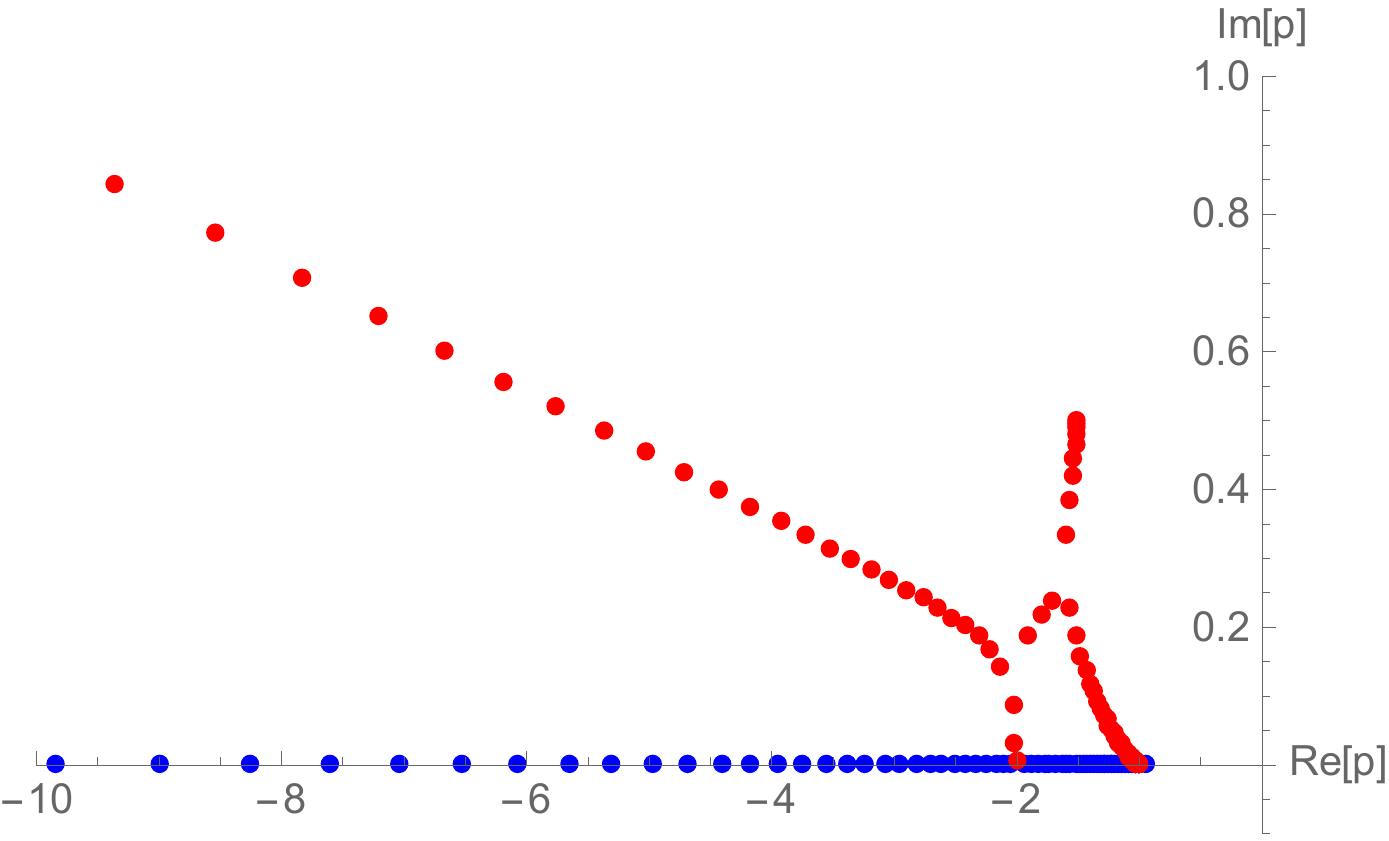}}
\caption{Pad\'e poles (blue) of the  Borel function $B(p)$ in (\ref{eq:probe1}), indicating a branch point at $p=-1$, and the Pad\'e poles of the Borel function with the addition of the ``probe singularity term'' in the text. The Pad\'e pole distribution is distorted but the genuine branch point singularity at $p=-2$ can now be seen, in addition to the probe singularity at $p=-\frac{3}{2}+\frac{i}{2}$. }
\label{fig:probe}
\end{figure}

\noindent{\bf Comments:}
\begin{itemize}[label=$\circ$]
\item 
A common problem of the Pad\'e-Borel approximation is that repeated Borel singularities may be hidden by the (unphysical) poles that Pad\'e generates to represent the branch cut associated with the leading singularity. The conformal map of the $\mathcal{PCB}$ approximation resolves this problem by separating the repeated singularities, as separated accumulation points on the unit circle in the conformally mapped $z$ plane \cite{Costin:2019xql,Costin:2020pcj}. For example, consider the Borel function
\begin{eqnarray}
B(p)=(1+p)^{-1/3} + \log(2+p)
\label{eq:probe1}
\end{eqnarray}
for which the  poles of a diagonal Pad\'e approximant are shown as blue dots in the first plot of Figure \ref{fig:hidden}. We clearly see the algebraic branch point at $p=-1$, but the logarithmic branch point at $p=-2$ is obscured by the Pad\'e poles attempting to represent the leading cut $p\in (-\infty, -1]$. However, after a conformal map (\ref{eq:cmap}) based on the leading singularity, we see that the second singularity at $p=-2$ is cleanly separated and identified, as the conformal map images $z=\pm i$. See the second plot in Figure \ref{fig:hidden}.

\item
Another simple method \cite{Costin:2020pcj}  to detect repeated singularities that are hidden by unphysical Pad\'e poles is based  on  the {\it potential theory} interpretation of Pad\'e poles as a configuration of charges that distribute themselves in such a way as to minimize the capacitance (relative to infinity) of the arcs along which Pad\'e places its poles (see Section \ref{sec:potential} and \cite{Stahl,Saff}). This means that we can introduce another ``test charge'' by adding to our truncated Borel transform function a singularity near the line of unphysical Pad\'e poles, near where we suspect a true singularity might be hidden. In a subsequent Pad\'e approximation the ``minimal capacitor'' will be distorted, but the genuine physical singularities do not move. 
For example, for the  Borel function in (\ref{eq:probe1}), 
we can add $B_{\rm probe}=(3/2-i/2+p)^{-1/7}$, which has a new singularity at $p=-\frac{3}{2}+\frac{i}{2}$. Then the Pad\'e pole distribution is distorted to the red dots in Figure \ref{fig:probe}. The second branch point at $p=-2$ is now clearly visible.

\end{itemize}

\section{Conclusions}

It is in general a challenging problem to extract physical information from a limited amount of perturbative information, which is often in the form of a finite number of terms of an expansion which is expected to be asymptotic. However, there are ways to combine Borel summation methods with suitable conformal and uniformizing maps in order to improve and optimize this process. Pad\'e approximants, and their physical interpretation in terms electrostatic potential theory, are particularly useful tools in this analysis. The technical challenge is to probe the singularity structure of the complex Borel plane, starting with only a finite number of coefficients in the expansion of the Borel transform, combined possibly with some information about the expected  global structure of the underlying Riemann surface. For realistic model examples, with the typical physical ``factorial times power'' rate of growth of coefficients (\ref{eq:bwl}), the gain in precision may be quantified, and is quite dramatic. In complicated physical systems for which only partial information is available,  these methods can also be used as non-rigorous (but extremely sensitive) exploratory tools to  refine approximate and conjectural results.

 \vspace{.3cm}
\noindent {\bf Acknowledgements} \\
This work is supported in part by the U.S. Department of Energy, Office of High Energy Physics, Award  DE-SC0010339 (GD).


\begin{thebibliography}{99}

 \bibitem{leguillou}
 J. C. Le Guillou and J. Zinn-Justin,
 {\it Large Order Behaviour of Perturbation Theory}, (North-Holland, 1999).

 
  \bibitem{ecalle} J. \'Ecalle, Fonctions Resurgentes, Publ. Math. Orsay 81, Universit\'e
   de Paris--Sud, Departement
   de Math\'ematique, Orsay, (1981).

  \bibitem{costin-book}
O. Costin,
{\it Asymptotics and Borel summability},
(Chapman and Hall/CRC, 2008).



\bibitem{Costin:2019xql} 
  O.~Costin and G.~V.~Dunne,
 ``Resurgent extrapolation: rebuilding a function from asymptotic data. Painlev\'e I,''
  J.\ Phys.\ A {\bf 52}, no. 44, 445205 (2019),
 \hhref{1904.11593}.
    
\bibitem{Costin:2020hwg}
O.~Costin and G.~V.~Dunne,
``Physical Resurgent Extrapolation,''
Phys. Lett. B \textbf{808}, 135627 (2020),
\hhref{2003.07451}.
 
\bibitem{Costin:2020pcj}
O.~Costin and G.~V.~Dunne,
``Uniformization and Constructive Analytic Continuation of Taylor Series,'' (2020), 
\hhref{2009.01962}.

    
    \bibitem{baker}
G. A. Baker, and P. Graves-Morris,
{\it Pad\'e Approximants}, (Cambridge University Press, 2009).



\bibitem{bateman} A. Erd\'elyi, Higher Transcendental Functions, The Bateman Manuscript Project, vol 1., New York--London (1953), \url{https://authors.library.caltech.edu/43491/}

  
\bibitem{Nehari} Z. Nehari, {\it Conformal Mapping}, Dover (1952).
 
\bibitem{zj} 
  J.~Zinn-Justin,
 {\it Quantum Field Theory and Critical Phenomena}, 
  Int.\ Ser.\ Monogr.\ Phys.\  {\bf 113}, 1 (2002).
  
    \bibitem{caliceti} 
  E.~Caliceti, M.~Meyer-Hermann, P.~Ribeca, A.~Surzhykov and U.~D.~Jentschura,
 ``From useful algorithms for slowly convergent series to physical predictions based on divergent perturbative expansions,''
  Phys.\ Rept.\  {\bf 446}, 1 (2007),
  \hhref{0707.1596}.
  
    
    \bibitem{caprini}
  I.~Caprini, J.~Fischer, G.~Abbas and B.~Ananthanarayan,
  ``Perturbative Expansions in QCD Improved by Conformal Mappings of the Borel Plane,''
  in {\it Perturbation Theory: Advances in Research and Applications}, (Nova
  Science Publishers, 2018),
 \hhref{1711.04445}.
 
 
\bibitem{bender}
C. M. Bender and S. A. Orszag,
{\it Advanced Mathematical Methods for Scientists and Engineers}, (Springer, 1999).
   
  \bibitem{Stahl} H. Stahl, 
  ``The Convergence of Pad\'e Approximants to Functions with Branch Points'', 
  J. Approx. Theory {\bf 91}, 139-204 (1997).

\bibitem{Saff} E. B. Saff, 
``Logarithmic Potential Theory with Applications to Approximation Theory'', 
Surveys in Approximation Theory, 5 (2010), 165-200, 
\hhref{1010.3760}.

\bibitem{szego}
G. Szeg\"o,
{\it Orthogonal Polynomials}, (American Mathematical Society, 1939);
U. Grenander and  G. Szeg\"o,
{\it Toeplitz forms and their applications}, (Univ. California Press, Berkeley, 1958).

\bibitem{Damanik-Simon} D. Damanik and B. Simon, 
``Jost functions and Jost solutions for Jacobi matrices, I. A necessary and sufficient condition for Szeg\"o 
asymptotics'', Invent. Math. {\bf 165}, 1-50 (2006).

    
  
\bibitem{b-g-w}  G. A. Baker, J. L. Gammel, and J. G. Wills, 
``An investigation of the applicability of the
Pad\'e approximant method'', J. Math. Anal. Appl. 2, 405-418. (1961).


    \bibitem{Lubinsky} 
    D. S. Lubinsky, 
    ``Rogers-Ramanujan and the Baker-Gammel-Wills (Pad\'e) conjecture'', 
    Annals of Math.  {\bf 157}, 847-889 (2003).

\bibitem{froissart}
M. Froissart,
``Approximation de Pad\'e: Application \`a la physique des particules \'el\'ementaires'',
\href{http://www.numdam.org/item?id=RCP25_1969__9__A2_0}{Les rencontres physiciens-math\'ematiciens de Strasbourg - RCP25, 1969, 
tome 9, 1-13 (1969).}

    
\bibitem{simon}
S. Graffi, V. Grecchi and B. Simon,
``Borel Summability: Application to the Anharmonic Oscillator'',
Physics Letters {\bf 32 B},  631-634 (1970).

\bibitem{marino}
M. Mari\~no,
{\it Instantons and Large N: An Introduction to Non-Perturbative Methods in Quantum Field Theory}, (Cambridge University Press, 2015).


  \bibitem{Yattselev} A. Aptekarev and M. L. Yattselev, 
  ``Pad\'e approximants for functions
    with branch points - strong asymptotics of Nuttall-Stahl polynomials'', 
    Acta Math. {\bf 215},  217-280 (2015).
    
\bibitem{abikoff} W. Abikoff, ``The Uniformization Theorem'', The American Mathematical Monthly, v. 88, No. 8, pp. 574-592 (1981).


\bibitem{schlag} W. Schlag, {\it A Course in Complex Analysis and Riemann Surfaces}, American Mathematical Society, Graduate Studies in Mathematics, vol. 154 (2014).     
    
    
   \bibitem{heller}
W.~Florkowski, M.~P.~Heller and M.~Spalinski,
``New theories of relativistic hydrodynamics in the LHC era,''
Rept. Prog. Phys. \textbf{81}, no.4, 046001 (2018),
\hhref{1707.02282}

\bibitem{serone} 
  M.~Serone, G.~Spada and G.~Villadoro,
   ``$\lambda \phi_2^4$ theory II. The broken phase beyond NNNN(NNNN)LO,''
  JHEP {\bf 1905}, 047 (2019),
  \hhref{1901.05023}.
  
  \bibitem{bertrand}
  C. Bertrand, S. Florens, O. Parcollet, and X. Waintal,
``Reconstructing Nonequilibrium Regimes of Quantum Many-Body Systems from the Analytical Structure of Perturbative Expansions'',
Phys. Rev. X {\bf 9}, 041008 (2019),
\hhref{1903.11646}.


\bibitem{rossi} 
  R.~Rossi, T.~Ohgoe, K.~Van Houcke and F.~Werner,
  ``Resummation of diagrammatic series with zero convergence radius for strongly correlated fermions,''
  Phys.\ Rev.\ Lett.\  {\bf 121}, no. 13, 130405 (2018),
\hhref{1802.07717}.


    \bibitem{hempel} 
    J. A. Hempel, 
    ``On the uniformization of the $n$-punctured sphere'', 
    Bull. London Math. Soc. {\bf 20}, 97-115 (1980).
    
\bibitem{kober} H. Kober, {\it Dictionary of Conformal Representations}, Dover (1957).  

 
   \bibitem{trefethen}
  A. Gopal,  L. N. Trefethen,
``Representation of conformal maps by rational functions'',
Numer. Math. {\bf 142}, 359-382 (2019), 
\hhref{1804.08127}.

 \bibitem{kuzmina}
 G.V. Kuz'mina, 
 ``Estimates for the transfinite diameter of a family of continua and covering theorems for univalent functions'',
 Proc. Steklov Inst. Math. {\bf 94}, 53-74 (1969).

 
  \bibitem{grassmann}
E. G. Grassmann and J. Rokne, 
``An explicit calculation of some sets of minimal capacity'',
SIAM J. Math. Anal. {\bf 6}, 242-249 (1975).


    
    \bibitem{Crowdy}
    D. G. Crowdy, 
    ``Schwarz-Christoffel mappings to multiply connected polygonal domains'',
Proc. Roy. Soc. A {\bf 461} (2005), 2653-2678.
  

  
\bibitem{Gukov:2016njj} 
  S.~Gukov, M.~Mari\~no and P.~Putrov,
 ``Resurgence in complex Chern-Simons theory,''
 \hhref{1605.07615}.


\bibitem{mckean}
H. P. McKean,
``Selberg’s Trace Formula as Applied to a Compact Riemann Surface'',
Commun. Pure Appl. Math. {\bf XXV}, 225-246 (1972).



  \end{thebibliography}
\end{document}